\documentclass[aps,reprint,floatfix,superscriptaddress,longbibliography]{revtex4-1}
\usepackage[english]{babel}
\usepackage[utf8]{inputenc}
\usepackage{epsfig}
\usepackage{amsmath}
\usepackage{units}
\usepackage[colorlinks]{hyperref}
\usepackage{enumerate}
\usepackage{color}

\renewcommand{\i}{\imath}
\newcommand{\komma}{,}

\newcommand{\ie}{{\it i.e. }}
\newcommand{\eg}{{\it e.g. }}

\newcommand{\tr}{\mathrm{Tr}}

\begin{document}

\title{Robustness of chiral edge modes in fractal-like-lattices below two dimensions:\\ A case study}
\author{Sonja Fischer}

\author{Michal van Hooft}

\author{Twan van der Meijden}

\author{Cristiane Morais Smith}

\author{Lars Fritz}

\author{Mikael Fremling}
\affiliation{Institute for Theoretical Physics and Center for Extreme Matter and Emergent Phenomena,
Utrecht University, Princetonplein 5, 3584 CC Utrecht, The Netherlands}

\begin{abstract}
One of the most prominent characteristics of two-dimensional Quantum Hall systems are chiral edge modes.
Their existence is a consequence of the bulk-boundary correspondence and their stability guarantees the quantization of the transverse conductance.
In this work, we study two microscopic models, the Hofstadter lattice model and an extended version of Haldane's Chern insulator.
Both models host Quantum Hall phases in two dimensions. We transfer them to lattice implementations of fractals with a dimension between one and two and study the existence and robustness of their edge states.
Our main observation is that, contrary to their two-dimensional counterpart,
there is no universal behavior of the edge modes in fractals. Instead,
their presence and stability critically depends on details of the models {\it and} the lattice realization of the fractal.
\end{abstract}

\maketitle

\section{Introduction}\label{sec:introduction}
The last decade has seen a lot of activity in the classification and realization of insulating electronic states that can be characterized by topological indices.
One of the milestones in the field has been the classification scheme known as the 'ten-fold way' or the 'periodic table' of non-interacting insulating states~\cite{kitaev2009periodic,qi2011topological,ryu2010topological}.
The two key ingredients in the scheme are the spatial dimension and symmetries, such as chirality, time-reversal, and parity.
For a certain combination of these, it allows to determine whether a topological index can be defined or not, and of which type it is. Importantly,
the periodic table does not require the existence of specific lattice symmetries,
implying that it also works in the presence of symmetry preserving disorder.

In more recent years, numerous extensions of this classification scheme have been introduced: crystalline topological insulators~\cite{fu2011topological},
non-Hermitian Hamiltonians~\cite{diehl2011topology},
driven non-equilibrium systems~\cite{lindner2011floquet},
or the so called higher order topological insulators~\cite{benalcazar2017quantized,schindler2018higher}, to name some of the prominent ones. 

An alternative path, pursued here, is to extend the classification scheme to allow for non-integer dimensions.
The 'tenfold-way' holds only for integer dimension, and an important question is what happens in between.
Or more directly, when precisely does a topological state cease to exist?  
Fractal structures can be defined as possessing non-integer Hausdorff dimension $d_f$~\cite{mcmullen1984hausdorff}.
Recently,
lattices with fractal features have been manufactured in the laboratory in a variety of ways.
This includes the use of molecular assembly~\cite{Shang2015,Tait2015,Nieckarz2016,Jiang2017,Sun2015,Zhang2015,Zhang2016}, templating, and co-assembly methods~\cite{Li2017a,Li2017b}.
Furthermore, fractal lattices were created by scanning tunneling microscope techniques~\cite{Kempkes2019} and with arrays of waveguides~\cite{Xu2020}.

Following the experimental developments, the theoretical interest in the physics of fractals has also been reviewed.
Recent works include investigations of the topology and the conductivity in Sierpinski carpets and gaskets~
\cite{
  Brzezinska2018,
  Fremling2020,
  vanVeen2017,vanVeen2016,Iliasov2019,Iliasov2019b,
  Yang2020b,
  Bouzerar2020,
  Sarangi2021}, 
but also studies of Floquet fractals~\cite{Yang2020a},
$p$-wave superconductors~\cite{Pai2019},
amorphous matter~\cite{Grushin2020} and even anyons in fractals~\cite{Manna2020}.

A special place in the 'periodic table' is taken by the class A. In two dimensions, this corresponds to
the class of the Integer Quantum Hall (IQH) effect, characterized by the absence of all of the above mentioned symmetries.
In the IQH in two dimensions, it is possible to define a bulk topological invariant, the Chern number,
which is an integer and directly related to the measurable transverse conductance $\sigma_{xy}$~\cite{TKNN1982}.
IQH systems are exceptionally robust due to their lack of symmetries.
They are not protected by symmetries but instead by the bulk gap.
This robustness of bulk properties directly translates to the edge properties.
The edges of the otherwise gapped IQH system host one dimensional chiral modes.
Each of these edge modes carries one quantum of conductance, $e^2/h$,
and admits ballistic transport, meaning that they are protected against backscattering from impurities.
As a consequence thereof, the transverse Hall conductance is quantized in units of $e^2/h$,
\ie $\sigma_{xy}= n e^2/h $. The integer $n$ is related to the Chern number of the bands below the Fermi energy,
but also counts the number of protected chiral edge modes in the finite system.

In this work, we study the existence and the robustness of edge modes when two-dimensional models of the IQH effect are transferred to fractal geometries with a Hausdorff dimension between one and two.
In practice, this means that we consider lattice implementations of fractals that are embedded in two dimensions and study microscopic tight-binding models belonging to the IQH class. Specifically, we study two microscopic models,
the Hofstadter model~\cite{Hofstadter1976} and a generalized version of the Haldane Chern insulator~\cite{Haldane1988} inspired by Ref.~\cite{Slager2013}.
Both models exhibit stable edge modes in two dimensions,
but they differ in one main aspect: the flux pattern in the Hofstadter model corresponds to a physical magnetic field, but the one in the Haldane Chern insulator model does not.

{\it Main result:}
We study both models on two different fractal lattice implementations,
the Sierpinski carpet and the Sierpinski gasket (or triangle). Our main finding is that 
there is no generic and universal stability of the edge modes, in contrast to their two dimensional counterpart. Instead,
the existence of current carrying edge modes strongly depends on microscopic details of both the fractals and its edge construction, as well as the models themselves. 

{\it Organisation of the paper:} The paper is organized as follows: We introduce the construction principle and the lattice implementations of the Sierpinski carpet and gasket/triangle in Sec.~\ref{sec:fractals}.
Subsequently, we discuss the two microscopic models and their key properties in Sec.~\ref{sec:models}.
We then present the computational method to calculate the transport properties in Sec.~\ref{sec:method}.
In Sec.~\ref{sec:carpet} and Sec.~\ref{sec:gasket}, we present results related to Hall transport on the respective carpet and gasket fractal lattice implementations and finish with a conclusion in Sec.~\ref{sec:conclusion}.
The more technical parts are relegated to appendices.

\section{Fractals and their lattice implementation}\label{sec:fractals}

The fractals considered in this paper are defined in the limit of an infinite iteration of a dilution scheme in the continuum that is specific to the fractal. On a lattice,
this iteration comes to a natural halt, once the lattice scale is reached.
This implies that all the fractals considered in this paper are only {\it approximate} fractals,
on a scale that is larger than the underlying lattice scale. 
We study two different implementations of fractals in this paper, the Sierpinski carpet and the Sierpinski gasket.
They differ in two important aspects: their fractal dimension and their edge connectivity.
Especially the latter plays a crucial role in our studies and is an important factor in determining the stability of the edge modes.

\subsection{The Sierpinski Carpet}\label{subsec:sierpinski}

\begin{figure}
    \includegraphics[width=.45\textwidth]{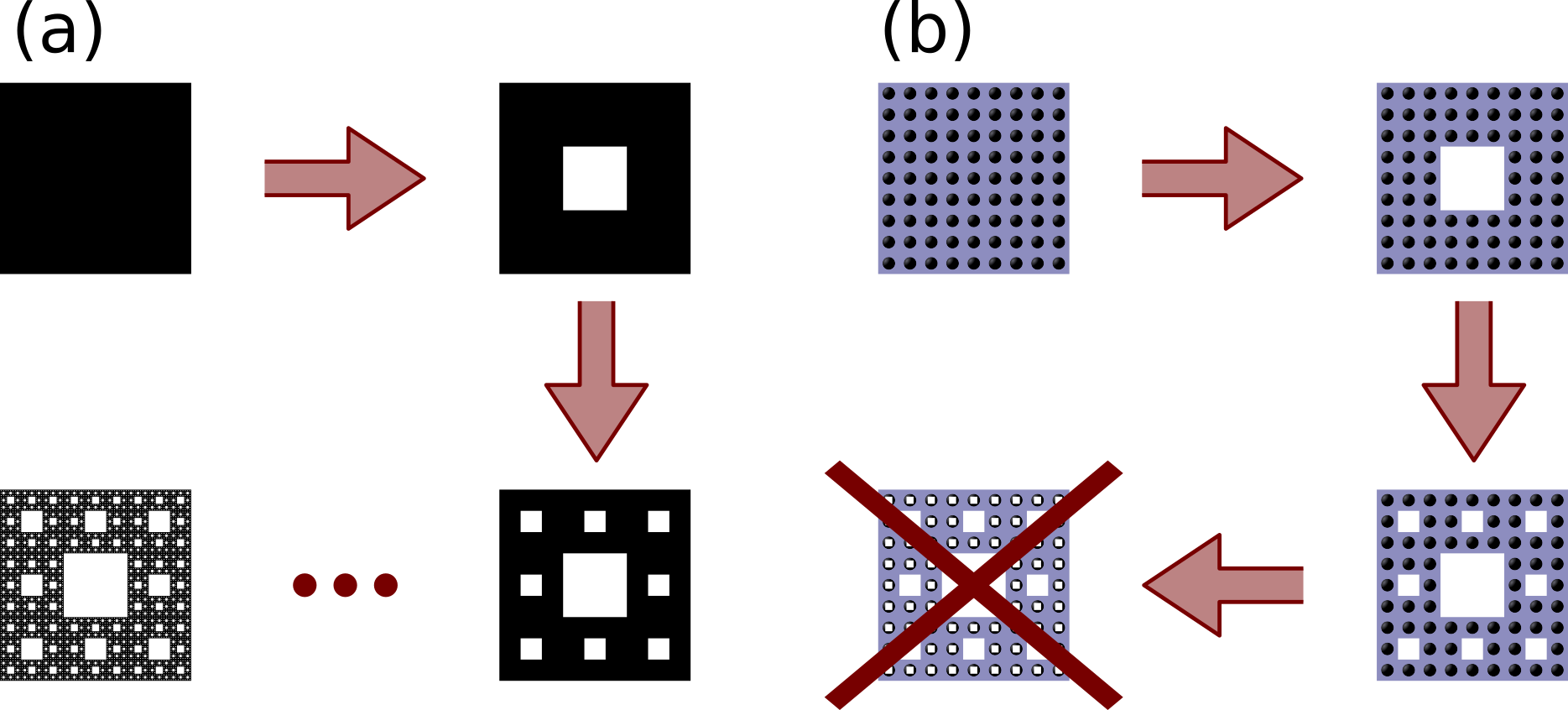}
    \caption{
      A graphical representation of the algorithm used to construct a Sierpinski carpet fractal.
      (a) the recursive procedure of cutting out the centers of every square. 
      (b)lattice implementation where a natural cut-off for the cutting procedure arises causing the recursion to stop.} \label{fig:sierpinskiconstruction}
\end{figure}

The Sierpinski carpet is a fractal that is constructed starting from a simple square through the iterative application of the cutting rule illustrated in the left-hand side of Fig.~\ref{fig:sierpinskiconstruction}.
One first divides a regular square into 9 squares of equal size.
Then the central square is removed and one is left with 8 smaller squares surrounding the empty central one.
The procedure is then repeated on each of the remaining squares.
The full fractal is obtained in the limit of an infinite repetition of this procedure.

\begin{figure}
    \centering
    \includegraphics[width=1.0\columnwidth]{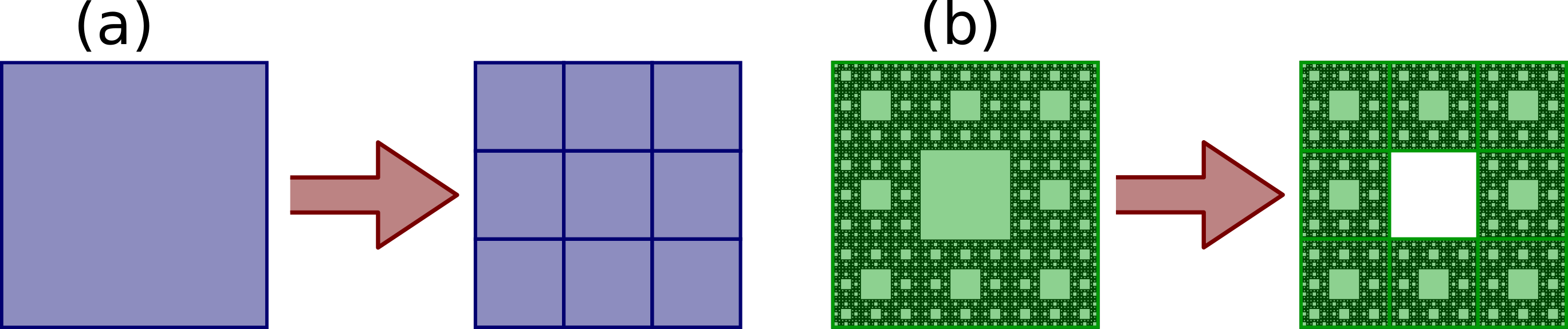}
    \caption{The scaling method for determining the Hausdorff dimension applied to the Sierpinski carpet (b) in contrast to a regular square (a).}\label{fig:fracdim}
\end{figure}

{\it Fractal Hausdorff dimension:} An illustrative and simple way to determine the Hausdorff dimension of the Sierpinski carpet is to investigate its scaling properties.
We consider the scaling exponent of the area under a single cutting procedure.
The square count is $8$ instead of $9$ because the centre was removed.
We then proceed to relate this rescaled area to the rescaled length of the outer square via the effective Hausdorff dimension $d_{\rm{H}}$
\begin{eqnarray}\label{eq:sierpinskidim_carpet}
    8&=&3^{d_{\rm{H}}} \quad {\rm{with}} \nonumber \\
    d_{\rm{H}}&=&\frac{\log 8}{\log 3}\approx 1.893\;.
\end{eqnarray}
A graphical illustration of the scaling procedure applied to the Sierpinski carpet is presented in Fig.~\ref{fig:fracdim}.

{\it Lattice implementation:} In order to connect the fractal with a tight-binding model, we have to define a lattice version of the above procedure.
For obvious reasons,	 the square lattice provides the most straightforward starting point for this.
On a finite lattice, the fractal always has a maximal depth,
\ie a maximum possible number of iterations for the cutting procedure.
This is illustrated in the right-hand side of Fig.~\ref{fig:sierpinskiconstruction},
where we regularize the lattice by placing sites on the centers of the squares in the Sierpinski carpet.
Technically, this construction should be named the {\it dual} Sierpinski carpet\cite{Melrose1983}, but for brevity we will simply refer to it as the Sierpinski carpet.

Since the maximal depth is related to the size of the lattice,
we can parameterize the side length of the square lattice as
\begin{eqnarray}
    \label{eq:SGdef}
    l=3^{G}a\;.
\end{eqnarray}
where $a$ is the lattice spacing.
The variable $G$ denotes the size generation, and is the maximal depth to which the fractal can be cut.

Given a lattice corresponding to a specific size generation $G$,
we are always free to \textit{not} cut to the maximal depth and instead stop at any earlier iteration of the cutting procedure. This proves to be a useful strategy for studying the edges.
Therefore, we introduce the quantity $F$, named fractal generation,
which denotes the number of cuts actually applied to our system. It is obvious that $F \leq G$ holds at all times.

\subsection{The Sierpinski Gasket}\label{subsec:sierpinskigasket}
\begin{figure}
\includegraphics[width=0.48\textwidth]{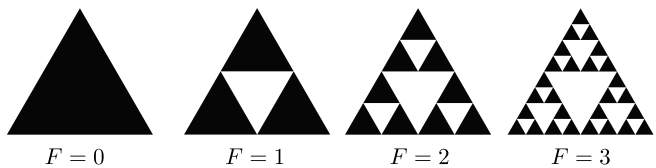}
\caption{Iterative construction of the Sierpinski gasket for fractal generation $F=0,..,3$.}
\label{fig:gasket}
\end{figure}

The Sierpinski gasket (SG) is constructed starting from an equilateral triangle in a very similar manner.
We first connect the centers of all three sides.
This forms another equilateral triangle sitting in the inside of the original triangle, but upside down.
In a next step we remove it, which leaves us with three triangles.
Afterwards,
the procedure is iterated on each of the remaining triangles, see Fig.~\ref{fig:gasket}.

{\it Fractal Hausdorff dimension:} The fractal Hausdorff dimension of the Sierpinski gasket, analogously to the discussion above,
can be obtained by considering the scaling exponent of the area with a single cutting procedure according to
\begin{eqnarray}\label{eq:sierpinskidim_triangle}
    3&=&2^{d_{\rm{H}}}\quad {\rm{with}}\nonumber \\
    d_{\rm{H}} &=&\frac{\log 3}{\log 2}\approx 1.585 \;.
\end{eqnarray}

\begin{figure}
  \centering
    \includegraphics[width=.94\linewidth]{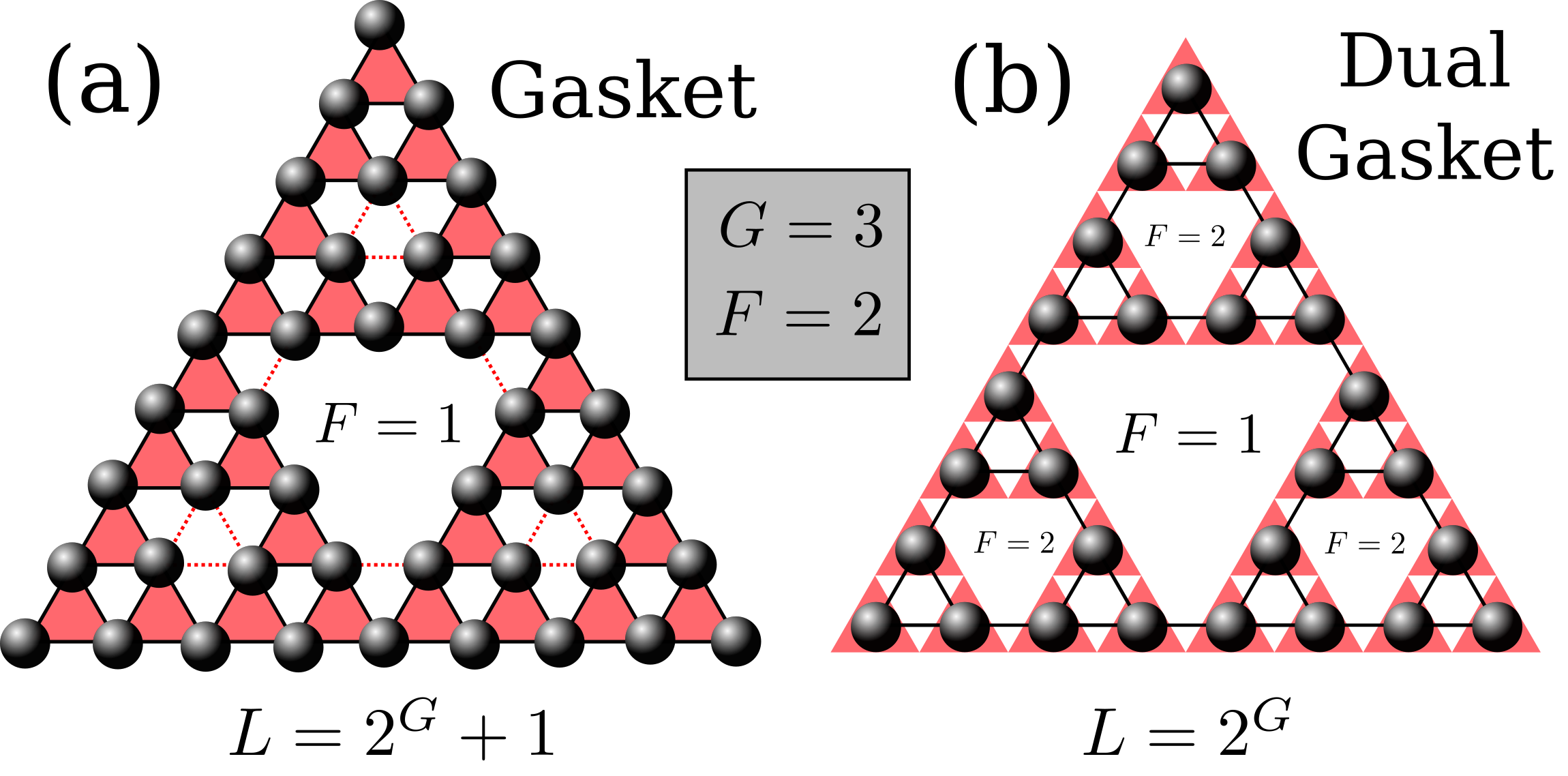}
  \caption{Two alternative regularization protocols. Protocol a),
Sierpinski Gasket (SG), places the sites on the vertices of the triangles.
Protocoll b), Dual Sierpinski Gasket (DSG), puts the sites on the centers of the triangles.
For nonzero $F$, in b) only sites are removed whereas in a) also bonds are cut, as indicated by red dashed lines.
In the thermodynamic limit, the two procedures yield the same Hausdorff dimension. }
  \label{fig:gasketlattice}
\end{figure}

{\it Lattice implementation:}

In the SG (left), the underlying triangular lattice is centered on the corners of the resulting triangles,
thus, under the cutting procedure, some sites needs to be removed, and some bonds need to be cut,
indicated by dashed red lines. Consequently, the remaining triangles share corners.
One could also build the dual SG which has the same Hausdorff dimension, but different connectivity and lacunarity.
The construction proceeds by assigning a lattice site to the centre of each triangle in the SG.

While both procedures produce the same thermodynamic limit,
the effect on the finite systems is very relevant, as the lacunarity is distinct.
In Sec.~\ref{sec:gasket}, we compare the two different lattice versions.
We find that the microscopic choice changes details of the physics, but not the overall conclusion that is drawn in this work. 

\begin{figure}
  \centering
  \includegraphics[width=0.45\textwidth]{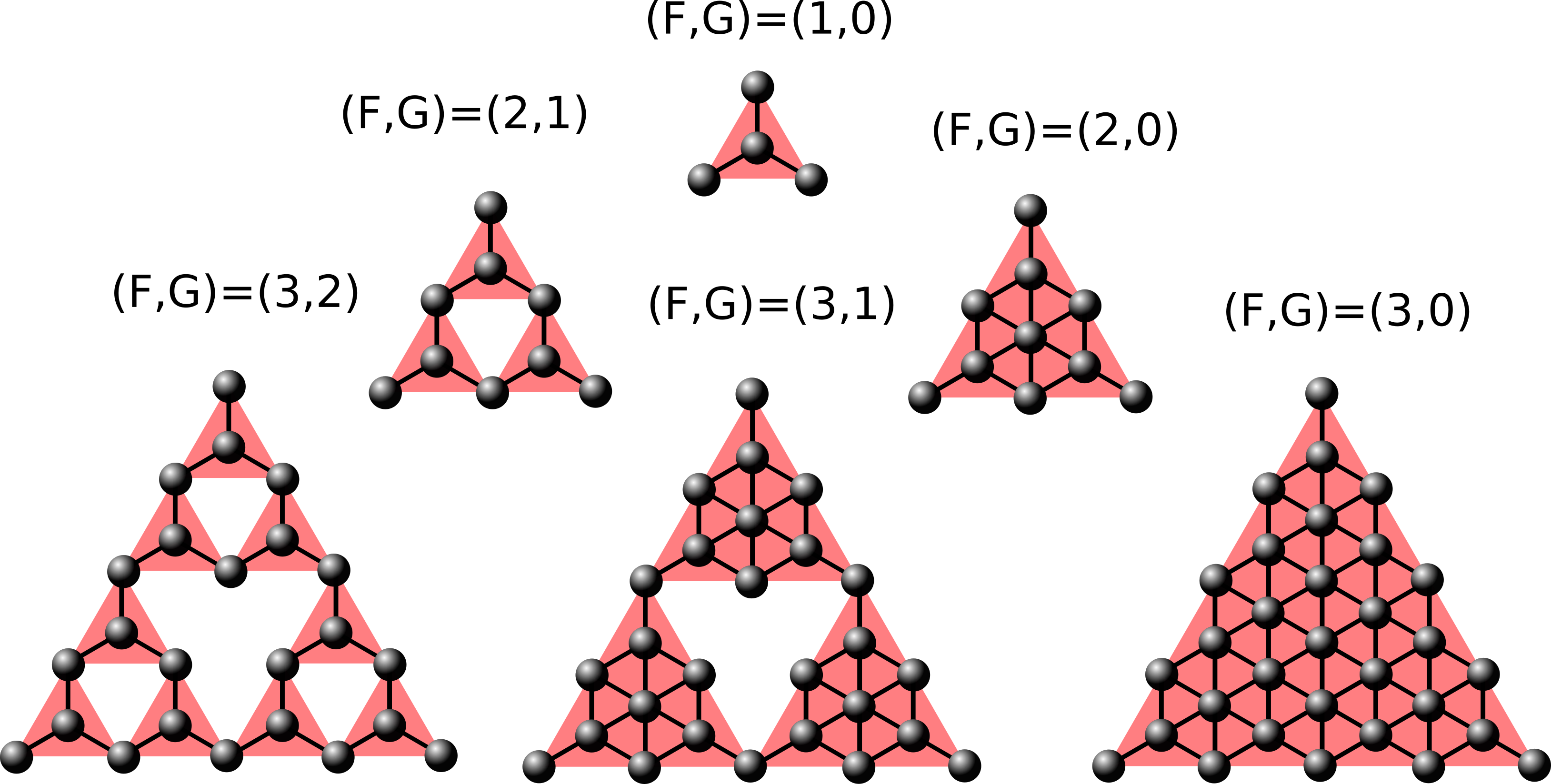}
  \caption{A third regularization scheme which places sites on both the corners and centers, which is not studied in this work.}
  \label{fig:GasketVariant}
\end{figure}

Note that these are by no means the only ways to implement the Sierpinski gasket on a triangular lattice.
One may for instance place the lattice points on both the corners and the edges\cite{Shang2015,Kempkes2019},
leading to a set of lattices displayed in Fig.~\ref{fig:GasketVariant}.
This lattice will, however, not be studied in this work. 

\section{The models}\label{sec:models}

We consider two prototypical models that both exhibit the IQH effect in two dimensions.
Specifically, we consider the Hofstadter lattice model and a version of Haldane's Chern insulator,
which we represent on a square lattice for computational convenience.

\subsection{The Hofstadter model}\label{sec:Hofstadter}

We only shortly review the salient features of the Hofstadter model and defer the interested reader to the existing literature for more details~\cite{Hofstadter1976,Beugeling2012,bernevig2013topological}.

\begin{figure}
\includegraphics[width=0.48\textwidth]{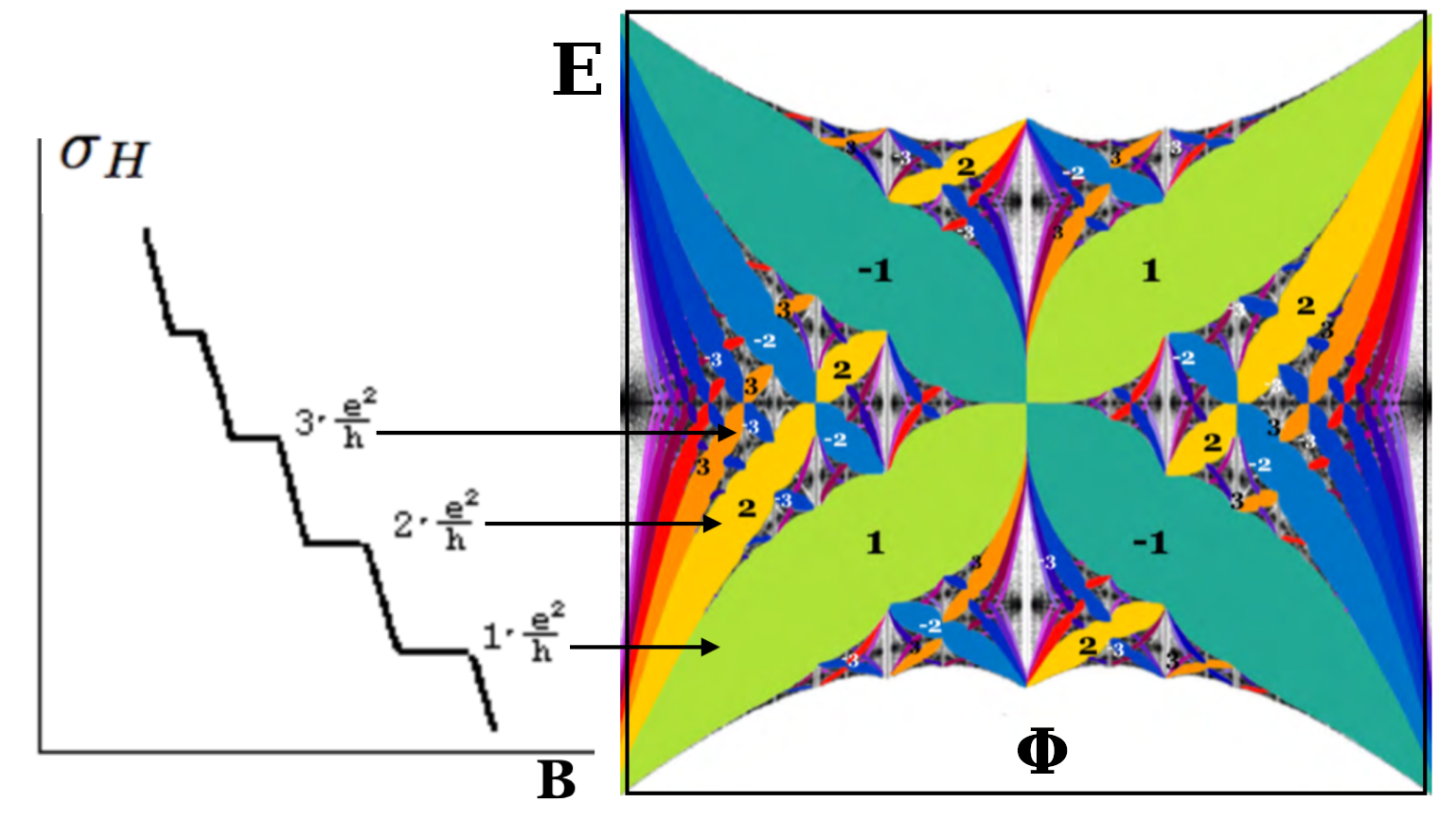}
\caption{Spectrum of the Hofstadter model. The gaps are labeled with the respective Chern numbers and the associated Hall conductivity.
  Courtesy of Refs.~\cite{Satija2016,Satija2018}.
}\label{fig:hofstadterphase}
\end{figure}

In its original version, the Hofstadter model is formulated on the square lattice~\cite{Hofstadter1976}.
Spinless electrons, created (annihilated) by $a_i^\dagger$ ($a_i^{\phantom{\dagger}}$),
hop on a square lattice under the influence of a homogeneous magnetic field that pierces all of its plaquettes.
The corresponding tight-binding Hamiltonian reads
\begin{equation}
  H_{\mathcal{L}}=-t\sum_{\left\langle i,j\right\rangle \in\mathcal{L}}
  \left( a_i^{\dagger} a_j^{\phantom{\dagger}} e^{\i A_{ij}}+\mathrm{h.c.} \right)\;,
  \label{eq:Frac_ham}
\end{equation}
where the magnetic field is implemented by means of the Peierls gauge connection $A_{ij}=e/h\int_{{\bf r}_i}^{{\bf r}_j} \vec A\cdot d\vec l$.
Here, $\mathcal{L}$ is the set of nearest neighbors with support on the lattice.
For concreteness, the magnetic field and the associated flux can be implemented in the Landau gauge $\vec A=B(y,0)$.
We parametrize its strength via $B=2\pi\Phi/a^2$,
where $\Phi$ is the magnetic flux piercing every plaquette.
For future reference, $\Phi_0=h/e$ is the quantum of flux, such that a flux of $\Phi_0$ can be trivially gauged away.
The Hofstadter model is most famous for its spectrum versus flux diagram,
which reveals the butterfly structure, Fig.~\ref{fig:hofstadterphase}. Additionally,
the gaps have been labeled by their respective Chern numbers, which have been calculated following the TKNN formula~\cite{TKNN1982}.
They are in one-to-one correspondence with the Hall conductivity. Importantly, by virtue of the bulk-boundary correspondence,
the Chern number is equal to the number of protected chiral edge modes.

\begin{figure}
  \begin{center}
    \includegraphics[width=0.48\textwidth]{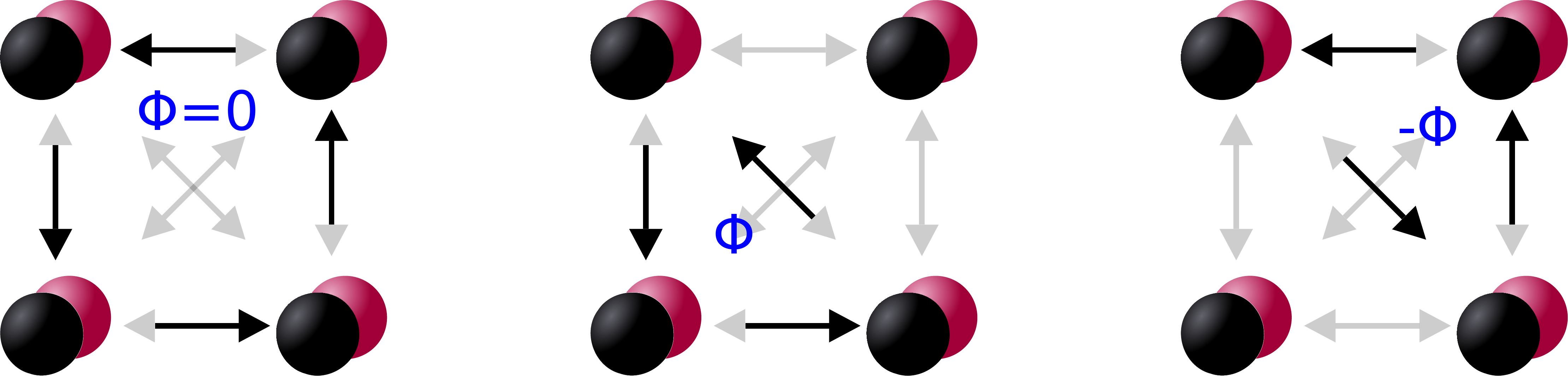}
  \end{center}
  \caption{Non-zero phases $\Phi$ can be acquired on closed loops that include next-nearest-neighbor hopping, as well as nearest-neighbor hopping.}
  \label{fig:fluxes}
\end{figure}

\subsection{A Haldane type Chern insulator}\label{sec:Chern}
The Haldane Chern insulator is a paradigmatic model that describes electrons hopping on the honeycomb lattice.
Its main feature is that it exhibits Quantum Hall physics without Landau levels,
as well as without a net magnetic flux~\cite{Haldane1988}. Here,
we study a variant of this model introduced in Ref.~\cite{Slager2013}. Contrary to the aforementioned paper,
we consider a spinless version which explicitly breaks time-reversal symmetry.
This is facilitated by complex next-nearest-neighbor hoppings in addition to nearest-neighbor hopping.
The model is again formulated on the square lattice and features two degrees of freedom per site, called $a$ and $b$.
The hopping pattern, shown in Fig.~\ref{fig:fluxes}, does not correspond to a net flux through the plane (the two degrees of freedom are graphically represented as black and red).

In real space the Hamiltonian reads
$H=H_{\circ}+H_{\uparrow}+H_{\downarrow}+H_{\rightarrow}+H_{\leftarrow}+H_{\nearrow}+H_{\searrow}+H_{\swarrow}+H_{\nwarrow}$,
and the arrow on each element of the total Hamiltonian indicates the direction in which an electron hops (and agrees with the sketch in Fig.~\ref{fig:fluxes}).
The term $H_{\circ}$ describes the on-site energy,
\[H_{\circ} = \sum_{i,j}\hat{\psi}^{\dagger}_{(i,j)}\left(M-8B\right)\sigma^z   \hat{\psi}_{(i,j)}\;,\]
where $\hat{\psi}$ is a two-component wave function, whose entries correspond to the two degrees of freedom,
 \ie $\hat{\psi}_i=\left(a_i,b_i \right)^T$ and $\sigma^{x,y,z}$ are the standard Pauli matrices acting in this subspace.
The nearest-neighbor hopping terms are given by
 \begin{widetext}
\begin{eqnarray} 
        H_{\uparrow}   &=&\sum_{i,j}\hat{\psi}^{\dagger}_{(i,j)}\left(B\sigma^z -\frac{1}{2\i}\sigma^y\right)\hat{\psi}_{(i,j+1)}    \;, \quad
        H_{\downarrow} =\sum_{i,j}\hat{\psi}^{\dagger}_{(i,j)}\left(B\sigma^z+\frac{1}{2\i}\sigma^y\right)   \hat{\psi}_{(i,j-1)}  \;, \nonumber \\
        H_{\rightarrow}&=&\sum_{i,j}\hat{\psi}^{\dagger}_{(i,j)}\left(B\sigma^z+\frac{1}{2\i}\sigma^x\right)   \hat{\psi}_{(i+1,j)}  \;, \quad
        H_{\leftarrow} =\sum_{i,j}\hat{\psi}^{\dagger}_{(i,j)}\left(B\sigma^z-\frac{1}{2\i}\sigma^x\right)   \hat{\psi}_{(i-1,j)}  \;,
\end{eqnarray}
and the next-nearest-neighbor hopping terms by
\begin{eqnarray} \label{eq:realspaceham}
  H_{\nearrow}   &=&\sum_{i,j}\hat{\psi}^{\dagger}_{(i,j)}\left(\tilde{B}\sigma^z+\frac{1}{4\i}\sigma^x + \frac{1}{4\i}\sigma^y\right)   \hat{\psi}_{(i+1,j+1)} \;, \quad
  H_{\searrow}   =\sum_{i,j}\hat{\psi}^{\dagger}_{(i,j)}\left(\tilde{B}\sigma^z-\frac{1}{4\i}\sigma^x + \frac{1}{4\i}\sigma^y\right)   \hat{\psi}_{(i+1,j-1)} \;, \; \nonumber \\
  H_{\swarrow}   &=&\sum_{i,j}\hat{\psi}^{\dagger}_{(i,j)}\left(\tilde{B}\sigma^z-\frac{1}{4\i}\sigma^x \frac{1}{4\i}\sigma^y\right)   \hat{\psi}_{(i-1,j-1)}\;, \quad
  H_{\nwarrow}   =\sum_{i,j}\hat{\psi}^{\dagger}_{(i,j)}\left(\tilde{B}\sigma^z\frac{1}{4\i}\sigma^x - \frac{1}{4\i}\sigma^y\right)   \hat{\psi}_{(i-1,j+1)} \;. 
\end{eqnarray}
 \end{widetext}

\begin{figure}
    \includegraphics[width=0.48\textwidth]{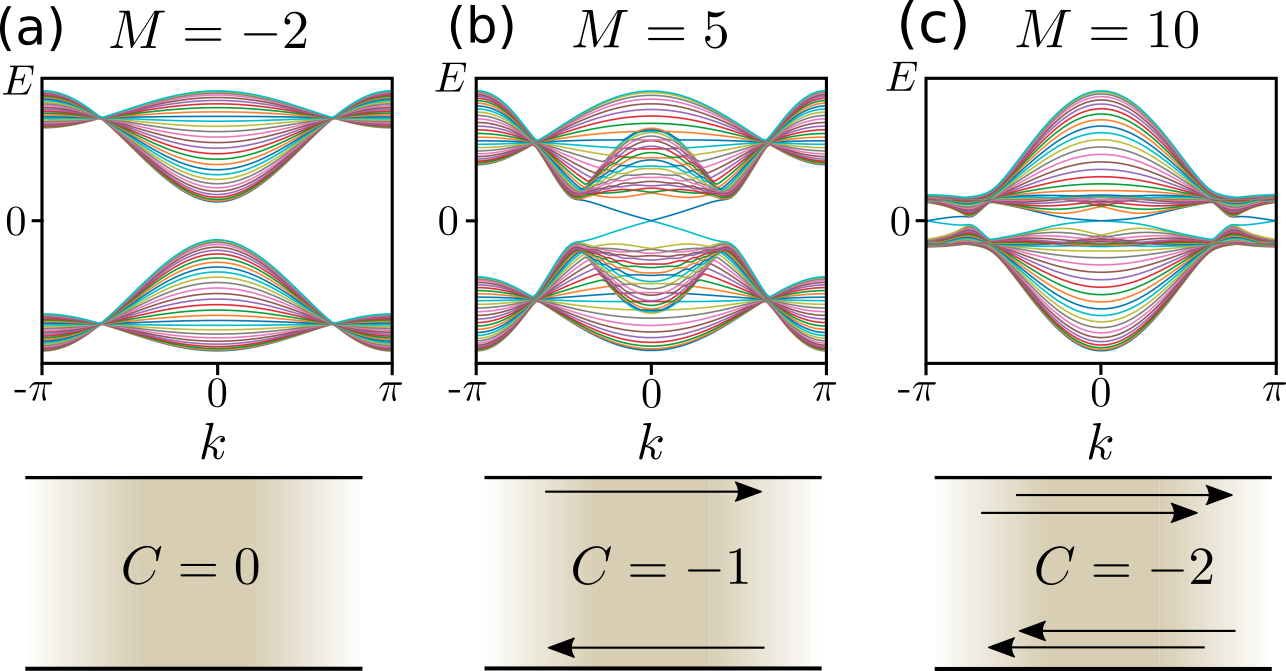}
    \caption{The band structure for a ribbon geometry with a width of 30 sites with two orbitals each.
The amount of in-gap boundary bands corresponds to the Chern number.
The counterpropagating edge states represented in (b) and (c) are at opposing ends of the cylinder.}
    \label{fig:ribbons}
\end{figure}

The index $i$ ($j$) describes the corresponding site's $x$-coordinate ($y$-coordinate).
The parameter $M$ is related to the difference in the local potentials between the $a$ and $b$ degrees of freedom,
while $B$ (${\tilde B}$) is proportional to the (next-)nearest-neighbor hopping between the degrees of freedom of the same type.
The absolute amplitude of both the nearest-neighbor and next-nearest-neighbor hoppings between different degrees of freedom are in the following set to  $B={\tilde B}=1$, leaving $M$ the only tunable parameter. 
While the net flux through each unit cell is zero, the model involves complex hopping parameters in a way to cause a quantum anomalous Hall effect (see Fig.~\ref{fig:fluxes}).

The Chern number $C$ can be computed in momentum space~\cite{bernevig2013topological}.
After imposing periodic boundary conditions and performing a Fourier transformation, the Hamiltonian can be written in the more compact form
\begin{eqnarray}\label{eq:Haldane}
H=\sum_{\bf{k}}\Psi^{\dagger}_{\bf{k}} \left[ {\vec{\sigma}} \cdot {\bf{d}}({\bf{k}})\right] \Psi^{\phantom{\dagger}}_{\bf{k}},
\end{eqnarray}
with
\begin{eqnarray}
  {\bf{d}}({\bf{k}})&=&\left(\begin{array}{c} \sin k_x+ \cos k_x \sin k_y \\ -\sin k_y +\cos k_y \sin k_x \\ f({\bf k})\end{array} \right),
\end{eqnarray}
and
\begin{eqnarray}
  f(\bf{k})&=&M-2B \left[ 2-\cos k_x -\cos k_y \right] \nonumber \\
  &&-4\tilde{B} \left[1- \cos k_x \cos k_y \right]\;.
\end{eqnarray}
The wave function reads $\Psi_{\bf{k}}= \left(a_{\bf{k}},b_{\bf{k}} \right)^T$,
and ${\vec{\sigma}}$ is the vector composed of the Pauli matrices,
\ie ${\vec{\sigma}}=(\sigma^x,\sigma^y,\sigma^z)$.

The Chern number for this model can be easily calculated according to $C=\frac{1}{4\pi}\int_{-\pi}^\pi dk_x dk_y \hat{d}\cdot\left ( \partial_{k_x}\hat{d}\times \partial_{k_y}\hat{d}\right) $, with $\hat{d}=\vec{d}/|\vec{d}|$.
We can identify three topologically distinct regions of the phase diagram, depending on the value of $M$, \ie
\begin{eqnarray}
    C=\left\{ \begin{array}{cc}
        -1\;, & 0<M<8 \;, \\
        -2 \; & 8<M<12 \; \\
        \phantom{-}0\; & \text{else}\;.\\
    \end{array} \right.
\end{eqnarray}
The number of protected chiral edge states is identical to the absolute value of $C$ (the sign of $C$ indicates the direction of propagation) for a given set of parameters.
We illustrate the connection between the Chern number and the number of edge modes on a cylinder geometry in Fig.~\ref{fig:ribbons}.

In the trivial phase ($C=0$), we see the band structure of an insulating system with no in-gap states.
In the $C=-1$ phase, there are two dispersive in-gap states,
one living on the upper boundary of the cylinder, while the other one lives on the lower boundary.
The states have opposite group velocities, \ie the slope of the eigenvalues have opposite signs, which means that the states travel in opposite directions.
For the $C=-2$ phase, we find two additional in-gap states.
An analysis of the group velocities and localization reveals
that edge states that have the same directionality are also localized on the same edge,
meaning they are chiral and there are two of them per boundary.

\begin{figure}
  \includegraphics[width=0.45\textwidth]{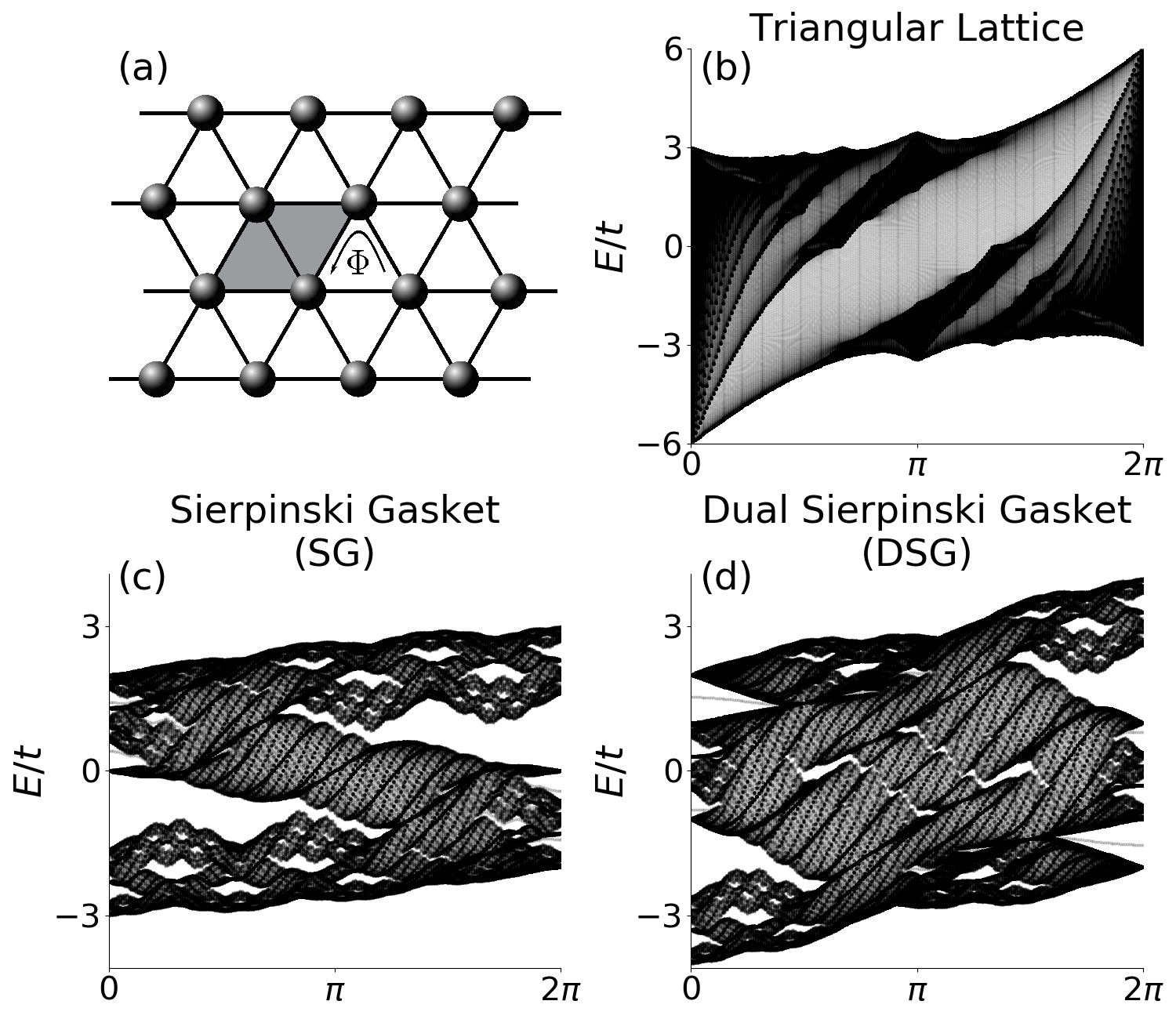}
\caption{(a) The triangular lattice and the flux $\Phi$ picked up around a triangle.
The gray area is the unit cell (or unit plaquette), which is $4\pi$ periodic in the flux $2\Phi$.
  (b) The spectrum of the Hofstadter model on a trianglular lattice as a function of the dimensionless flux $\Phi$.
The spectrum of the Sierpinski gasket strongly depends on whether it is regularized as a gasket (c) or a dual gasket (d). 
}\label{fig:butterflytriangle}
\end{figure}

\subsection{The Hofstadter model on the triangular lattice}\label{sec:Hofstadtertriangle}

This model is formally equivalent to the one discussed in Sec.~\ref{sec:Hofstadter}, in that it is described by the same Hamiltonian.
The main difference, however, is that the smallest closed loop is not given by a plaquette but by a triangle, and the choice of gauge is slightly more involved.
The smaller loop means that the system is not $2\pi$ periodic in the phase picked up per plaquette,
but is $4\pi$ periodic, see Fig.~\ref{fig:butterflytriangle} (a).

The spectrum of the 'butterfly' as a function of the flux $\Phi$ through a lattice is shown in Fig.~\ref{fig:butterflytriangle} (b).
While it looks different than the one in the square lattice, it has very similar features.
It also shows large regions with bulk gaps, where the system is a IQH system characterized by an integer quantized Hall conductance.
For all practical purposes, the system behaves similarly to the more conventional square lattice version.
The main difference will be where we put the leads, as detailed in Sec.~\ref{sec:gasket}.
In Figs.~\ref{fig:butterflytriangle} (c) and (d), we contrast the spectra of the SG and the dual SG both from the triangular lattice, as well as from each other.
The main observation is that the spectra of SG and dual SG are markedly different as a consequence of the two distinct cutting procedures.

\section{The method}\label{sec:method}

\begin{figure}
  \includegraphics[width=.40\textwidth]{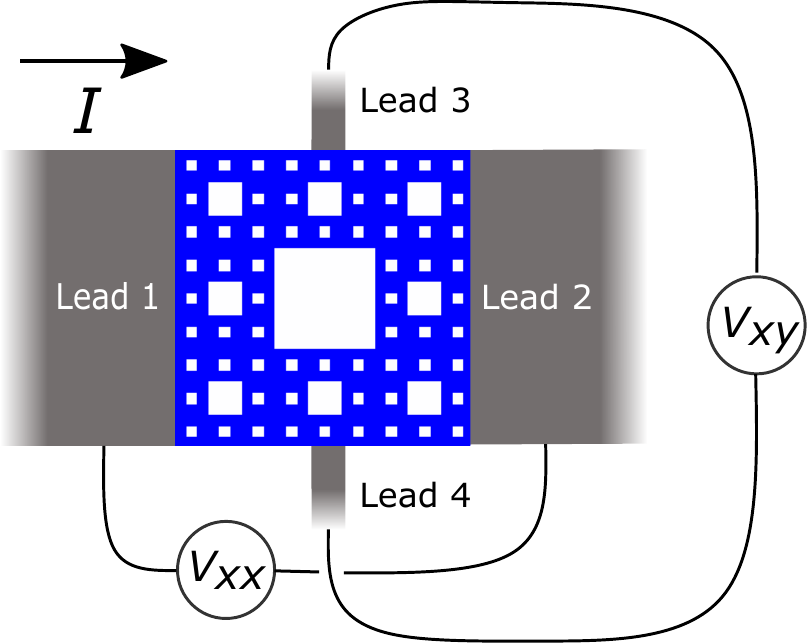}
  \caption{Four terminal setup, which allows to access both the diagonal and the transverse voltage drop.}
  \label{fig:fourterminal}
\end{figure}

We use two complementary methods to study the properties of the edge states: eigenstate spectroscopy and the non-equilibrium Green function technique.
The former one is straightforward, so we do not comment on it any further. The latter,
on the level it is used here, is equivalent to the Landauer-B\"uttiker approach.
We implement it numerically and in order to achieve larger system sizes, we combine it with the recursive Green function method.
The setup we study for square based geometries is shown in Fig.~\ref{fig:fourterminal}.
It also applies to the triangular systems, although with slightly modified lead locations.
Four leads are connected to the central scattering region ($S$) that hosts the fractal.
The leads can in principle sit at different temperatures $T_a$ and at different chemical potentials $\mu_a$ ($a=1,2,3,4$).

The main advantage of this setup is that it allows to study both the diagonal voltage drop $V_{xx}$ and the Hall or transverse voltage drop $V_{xy}$, which provide complementary insights. 

\begin{figure}
  \includegraphics[width=0.48\textwidth]{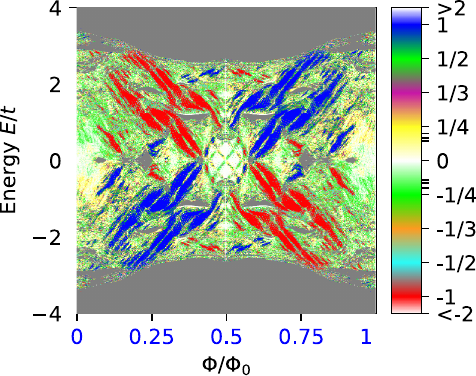}
  \caption{The Hofstadter model phase diagram, in terms of $\rho_{x,y}$ for a full depth fractal of a generation 5 Sierpinski carpet.
Picture from Ref.~\cite{Fremling2020}.}\label{fig:Hall_Butterfly}
\end{figure}

\begin{figure*}[tb]
  \begin{centering}
    \includegraphics[width=0.98\textwidth]{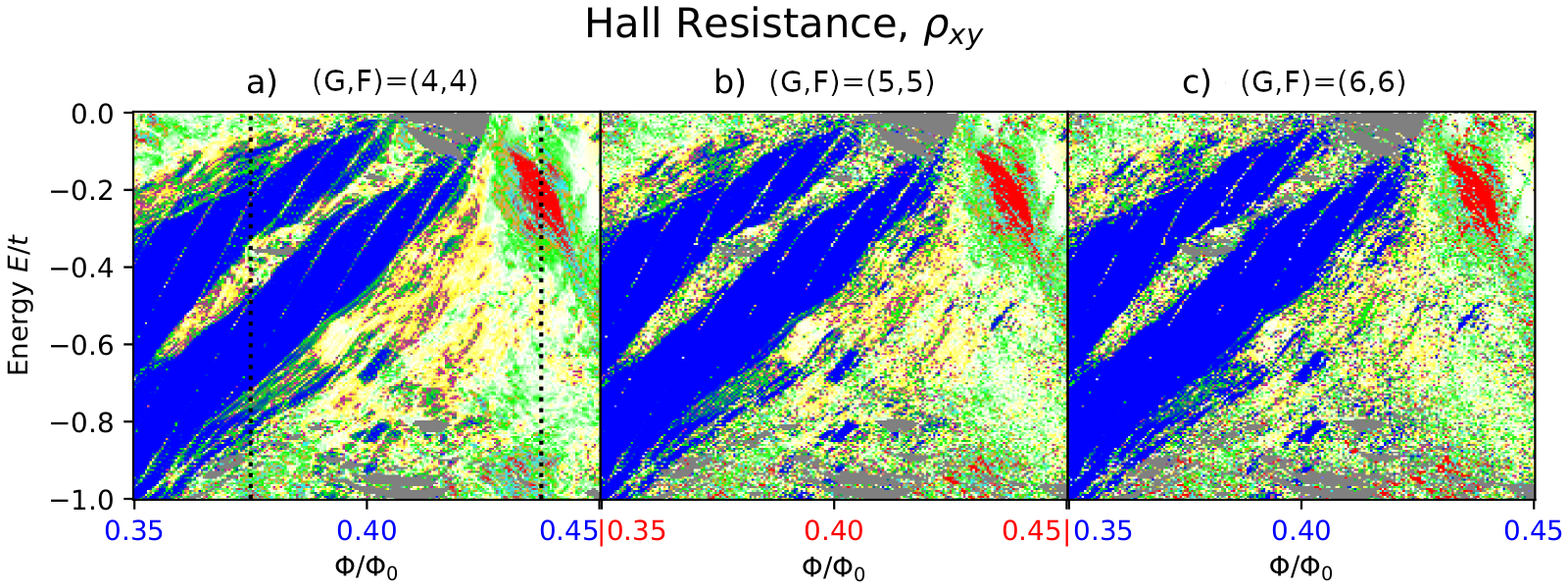}\\
    \includegraphics[width=0.98\textwidth]{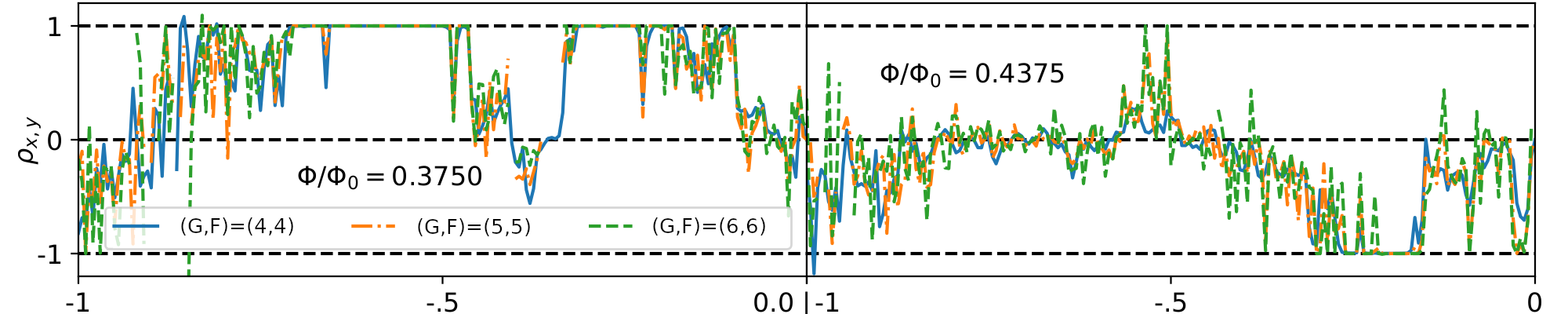}
    \par\end{centering}
    \caption{Zoom in on the Hall resistance for three consecutive generations $(4,4)$, $(5,5)$ and $(6,6)$.
      The large scale structure of the edge modes is already present in generation $(4,4)$.
      The difference between the various panels is the amount of high frequency variations,
as seen in the two cuts at fixed $\Phi/\Phi_0$ in the lower panel.
      The color scheme is the same as in Fig.~\ref{fig:Hall_Butterfly}.
      Picture from Ref.~\cite{Fremling2020}.
      \label{fig:Hall_Butterfly_Zoom_in}}
\end{figure*}

\begin{figure*}[t]
  \includegraphics[width=0.98\textwidth]{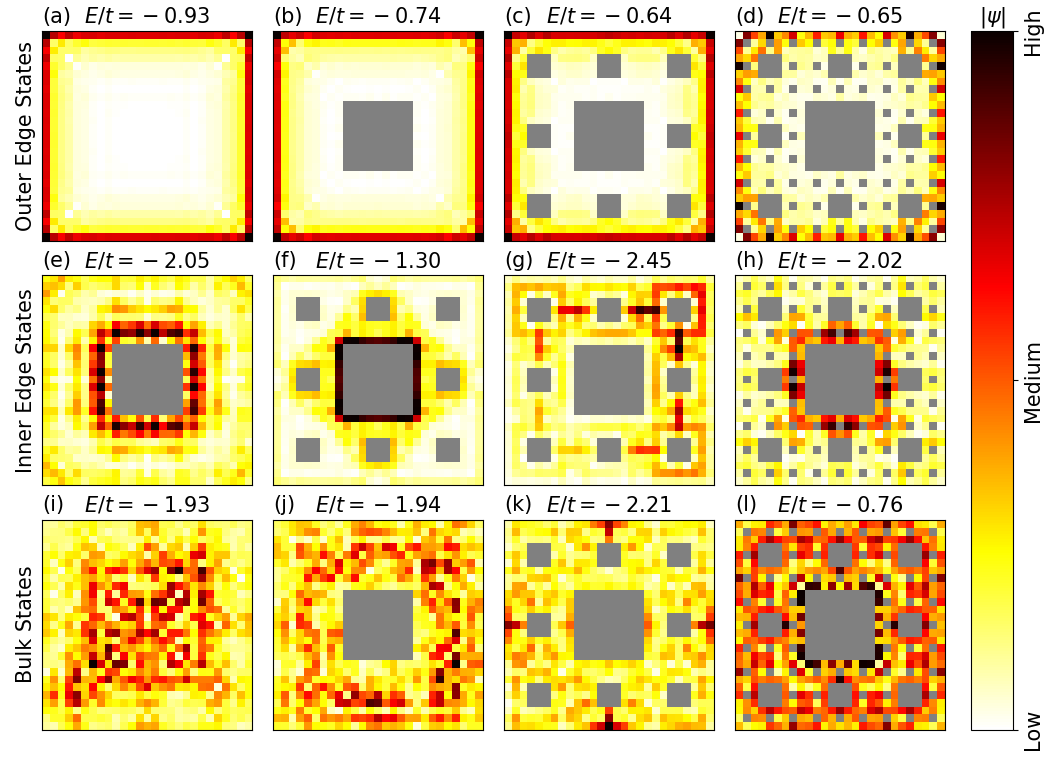}
  \caption{Upper Row: The structure of edge states at $\Phi/\Phi_0=0.375$ for all depths of a fractal of generation 3.
    The states chosen lie in an energy range with quantized $\rho_{xy}$ in Fig.~\ref{fig:Hall_Butterfly_Zoom_in}.
    Note how all the states are essentially localized to the outermost site,
    which enables their stability also at full depth of the fractal.\\ Middle Row: Inner edge states at the same value of flux.
These states behave like edge states localized around the holes of the Sierpinski Carpet.\\
    Lower Row: Bulk states outside of the zoom in Fig.~\ref{fig:Hall_Butterfly_Zoom_in}.
    In all figures random uniform onsite disorder in the range $\epsilon\in[-1,1]\cdot 0.01t$ is added to break residual lattice symmetries.
  }\label{fig:Edge_states}
\end{figure*}

Following a standard Keldysh Green function treatment, one can derive the following expression for the current into lead $a$:
\begin{eqnarray}
I_{a}=\frac{e}{h}\sum_{b=1}^4\int \text{d} \omega \, T_{a,b}(\omega)\left[f_a(\omega)-f_b(\omega)\right]\;,
\end{eqnarray}
where
\begin{eqnarray}\label{eq:transmission}
T_{a,b}(\omega)=\tr\left[\Gamma_a(\omega) G^R_{a,b}(\omega)\Gamma_b(\omega) G^A_{a,b}(\omega)\right]
\end{eqnarray}	
is the transmission function between leads $a$ and $b$. The trace in Eq.~\eqref{eq:transmission} extends over all internal indices,
such as for instance orbitals and the sites along the length of the interface.
$G^{R(A)}_{a,b}$ and $G^{R(A)}_{b,a}$ are the retarded (advanced) Green functions describing propagation from lead $a$ to lead $b$ and vice versa (these are generically large matrices that have to be handled numerically).
The broadening function $\Gamma_{a}(\omega)$ describes the hybridization function of lead $a$ with the scattering region $S$.
It is given by
\begin{eqnarray}
    \Gamma_{a}&=&\i \left(\Sigma^R_a - \Sigma^A_a\right)\;,\nonumber \\
    \Sigma_{a}&=& V_{S\rightarrow a}G_{a}V_{a\rightarrow S}\;,\label{eq:Broadening}
\end{eqnarray}
where $G_a$ is the lead Green function, meaning the Green function of the decoupled lead at the interface. This quantity is also calculated numerically.

Furthermore, $f_{a}$ denotes the Fermi-Dirac distribution within lead $a$,
\begin{eqnarray}
    f_{a}(\omega)=\left \{1+\exp\left [ -\beta_{a}\left(\omega-\mu_{a}\right)\right ] \right \}^{-1}\komma
\end{eqnarray}
where $\beta_{a}=(k_BT_{a})^{-1}$ describes the inverse temperatures and $\mu_{a}$ is the chemical potential of the respective lead $a$.
In the following analysis, we always consider zero temperature in the leads,
implying that the distribution function is the step function. Consequently, the current into lead $a$ reads
\begin{eqnarray}
  I_a&=& \frac{e}{h}\sum_{b=1}^4\int_{\mu_a}^{\mu_b} \text{d} \omega \, T_{a,b}(\omega)\\
  &=&\frac{e}{h}\sum_{b=1}^4(V_b-V_a)T_{a,b}(\mu)\,
\end{eqnarray}
where in the last line we assume $\mu_{a(b)}=\mu+V_{a(b)}$ and $T_{a,b}(\omega)\approx T_{a,b}(\mu)$.
In this paper, we are mostly concerned with calculating the transverse resistance and, to a lesser extent, the longitudinal one.
The pattern of currents that allows us to access both is given by ${I_1,I_2,I_3,I_4}={I_0,-I_0,0,0}$.
If we ground \eg lead no. $4$, \ie $V_{4}=0$, we can determine all the potentials and the Hall resistivity is given by 
\begin{eqnarray}
	\rho_{xy}=\frac{V_3}{I_0}\;.
\end{eqnarray}

The main advantage of the above formulation in terms of Green functions is that we are able to express the transmission in terms of only a very limited number of elements of it.
This allows to use a very efficient numerical technique, the recursive Green function method.
The algorithm has the benefit of reducing computation time significantly.
Instead of inverting the whole central scattering region $S$, one can do it iteratively, slice by slice.
However, it needs to be adapted to accommodate the presence of more than two leads.
Technical details on the Recursive Green function formalism and slicing procedure are given in Appendices \ref{app:recgf} and \ref{app:slicing} for the interested reader.

\section{Edge states at the boundary of the Sierpinksi carpet fractal}\label{sec:carpet}

In this section, we discuss the physics of the edge states on the Sierpinski carpet of both the Hofstadter and the Chern insulator model.
Our main finding is that the stability of the edge states of both two dimensional models differ significantly upon rendering the lattice fractal.  

In a strictly two dimensional setting, the Hall conductivity $\sigma_{xy}$ is quantized according to $\sigma_{xy}=Ce^2/h$,
where $C$ is the Chern number of the underlying electronic structure.
Alternatively, the transverse conductivity can be considered from the point of view of chiral edge modes in the system,
following the bulk-boundary correspondence.
It is important that the bulk itself is gapped.
Then, the Hall conductivity is quantized as $|\sigma_{xy}|=n e^2/h$ where $n$ is the number of protected chiral edge modes.
Since the Chern number is only properly defined and quantized for gapped systems in two dimensions (or more generally,
in even dimensions), we will henceforth not talk about the Chern number but only about $\sigma_{x,y}$ (and $\rho_{x,y}$), since this quantity is always well defined.

\begin{figure}
  \begin{center}
    \includegraphics[width=0.50\textwidth]{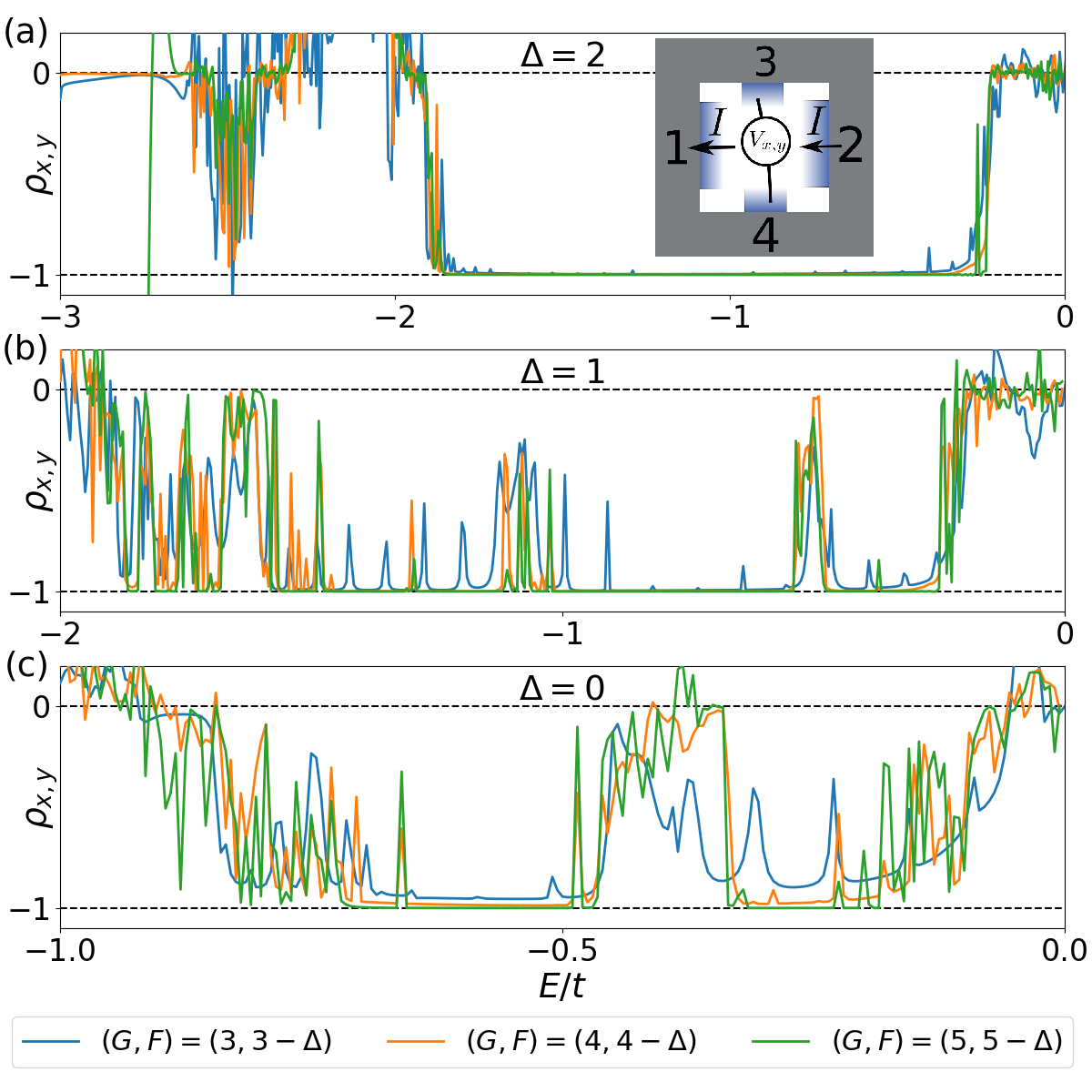}
  \end{center}
  \caption{Hall resistance for leads placed on the interior of the largest hole in the Sierpinski carpet,
at a fixed flux $\Phi/\Phi_0=0.375$, and fractal distance $\Delta=2,1,0$.
    It should be evident that one can detect the signature of interior edge modes even at the largest depths, $\Delta=0$,
just like for the outer edge modes in Figs.~\ref{fig:Hall_Butterfly_Zoom_in} and \ref{fig:Hall_Butterfly}.
  }
  \label{fig:InnerCarpet}
\end{figure}

\subsection{The Hofstadter model on a Sierpinski carpet}

In this part, we investigate the generalization of the Hofstadter model to a fractal geometry.
Some of the results have already been published elsewhere~\cite{Fremling2020}, and we only give a very condensed version here.
The main results are summarized in Fig.~\ref{fig:Hall_Butterfly} and Fig.~\ref{fig:Hall_Butterfly_Zoom_in}.
Similar results have also been obtained in Ref.~\onlinecite{Iliasov2019b} using the Kubo-Bastin   formula   for Hall  conductivity.

Fig.~\ref{fig:Hall_Butterfly} shows the Hall resistivity for a generation 5 fractal at full depth, $G=F$.
The $x$-axis denotes the flux value per plaquette of the parent square lattice, in units of the flux quantum.
On the $y$-axis, we show the energy in units of the elementary hopping introduced in Eq.~\eqref{eq:Frac_ham}.
The color scale shows the dimensionless transverse resistivity $\rho_{xy}$ in units of $h/e^2$.
Compared to the square lattice, Fig.~\ref{fig:hofstadterphase},
the number of regions with perfect quantization is massively reduced.

It was found that the vital quantity that determined the maximum number of edge modes was not $G$ or $F$, but the difference between them, the 'fractal distance'
\begin{equation}
    \label{eq:Deltadef}
    \Delta=G-F \;.
\end{equation}
We will see that this quantity plays a crucial role also in the systems that we consider in this work.
This becomes more apparent in Fig.~\ref{fig:Hall_Butterfly_Zoom_in},
which shows a zoom-in into the upper panel of Fig.~\ref{fig:Hall_Butterfly}.
From left to right, we increase the fractal generation (all at full depth, $\Delta=0$), making sure that the results have converged.
In the lower panel, we see a corresponding cut at fixed flux, again for different generations.

For some fluxes  we find plateaus, whereas for others no quantization is visible.
This suggests that generically the quantum Hall physics of two dimensions is unstable to modifying the dimension.
From the point of view of Chern numbers and the periodic table,
this is not unexpected~\cite{kitaev2009periodic,qi2011topological,ryu2010topological}.
It is not obvious how to define Chern numbers if the dimension is not two,
and therefore one could expect that for fractals the quantization, in general, is gone.

The specific problem with fractals is that even if the embedding space is two-dimensional the system does not possess any finite period, which prevents the formation of a two-dimensional Brillouin zone.
One can artificially impose periodic boundary conditions on any finite-size fractal.
However, these ``periodic'' systems would have just as many bands as there are sites in the fractal, and this number would grow with system size, preventing the number of bands from being a well defined quantity in the thermodynamic limit.

Alternative methods are currently being developed to describe the topology in systems lacking translation invariance.
These include calculations in real-space\cite{Kitaev2006}, perpendicular space\cite{Rai2021} or using non-commutative geometry techniques\cite{Bellissard1994,Prodan2011,Prodan2017}.
The real-space calculations on Sierpinski fractals have been attempted in Refs.~\onlinecite{Brzezinska2018} and \onlinecite{Fremling2020}, but as far as we are aware, these do not guarantee quantization.

Since the quantization of the Hall conductivity is also related to the existence of chiral edge states (assuming there is a bulk gap), this begs the question of what happens to them.

A priori nothing forbids the existence of edge states in a fractal between one and two dimensions.
In order to analyze this, we consider the modes that correspond to states at the Fermi level in one of the plateaus shown in Fig.~\ref{fig:Hall_Butterfly_Zoom_in}.
We choose a flux of $\Phi/\Phi_0=0.375$ and identify edge states at all depths of a fractal of generation $4$, see Fig.~\ref{fig:Edge_states} (upper row). 
Given that edge modes can survive on the exterior edges,
it is natural to ask whether the same is true also on the interior edges,
which are formed by the holes that are cut away from the fractal.
Indeed, as shown in the middle panel of Fig.~\ref{fig:Edge_states},
one may identify states that are localized on the inner edges as well (note that we have added weak disorder in all plots to remove accidental degeneracies due to lattice symmetries).
These states are also identified in Fig.~\ref{fig:InnerCarpet} via a Hall resistance measurement involving leads placed on the inner edges instead of the outer ones. 
In the figures, the leads are placed in the central, $F=1$, ``hole'' and $\rho_{x,y}$ is computed for $\Delta=0,1,2$ and $G=3,4,5$.
We find, again, that $\Delta$ determines whether or not edge modes are stable on the inner edge, just like on the outer ones. 
The key to understanding the stability and the associated quantization plateau is that for some energies the wave function is localized on the outermost boundary sites.
This is a fine tuned situation and it renders the system less stable against adding disorder, as some of us showed in Ref.~\cite{Fremling2020}.
For reference, a number of typical bulk states are shown in the lower panel of Fig.~\ref{fig:Edge_states} for the same cutting depths.

\subsection{The Haldane Chern insulator on a Sierpinski carpet}

We now turn our attention to the Haldane  model defined in Eq.~\eqref{eq:Haldane}.
The setup we consider is of the type shown in Fig.~\ref{fig:fourterminal}.
Our starting point is the square lattice in the scattering region $S$, to which we then 
apply the cutting algorithm to generate the Sierpinski carpet. 
We first investigate the Hall resistance for signatures of quantization, in the sense of the IQH effect. 

\begin{figure*}
  \begin{center}
  \includegraphics[width=0.98\textwidth]{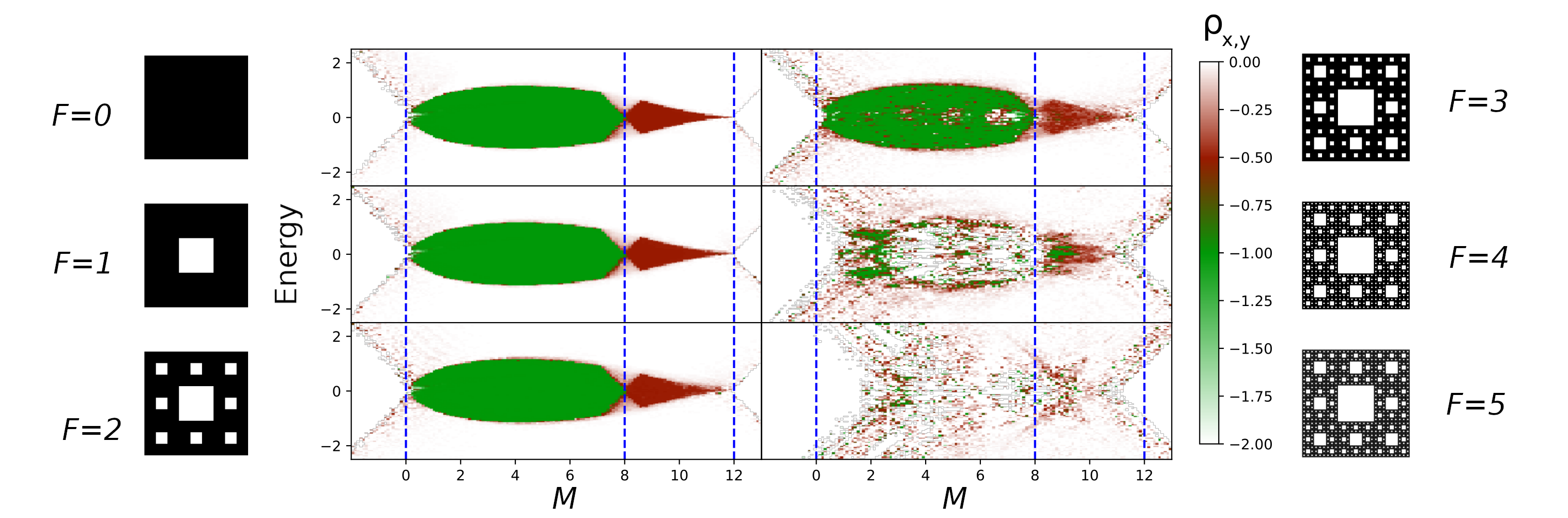}
  \end{center}
  \caption{A comparison of Hall resistance at all possible iterations of the cutting procedure for a system with size $G=5$.
    While edge modes survive the first few cuts, they break down for $F\geq3$.}
  \label{fig:gen5all}
\end{figure*}

Starting from a two-dimensional system of size $G=5$, we compute the Hall resistance for different cutting depths, \ie for $F$-values between $0$ and $5$.
The results are shown in Fig.~\ref{fig:gen5all}.

\begin{figure*}[htb]
  \begin{center}
  \includegraphics[width=.98\textwidth]{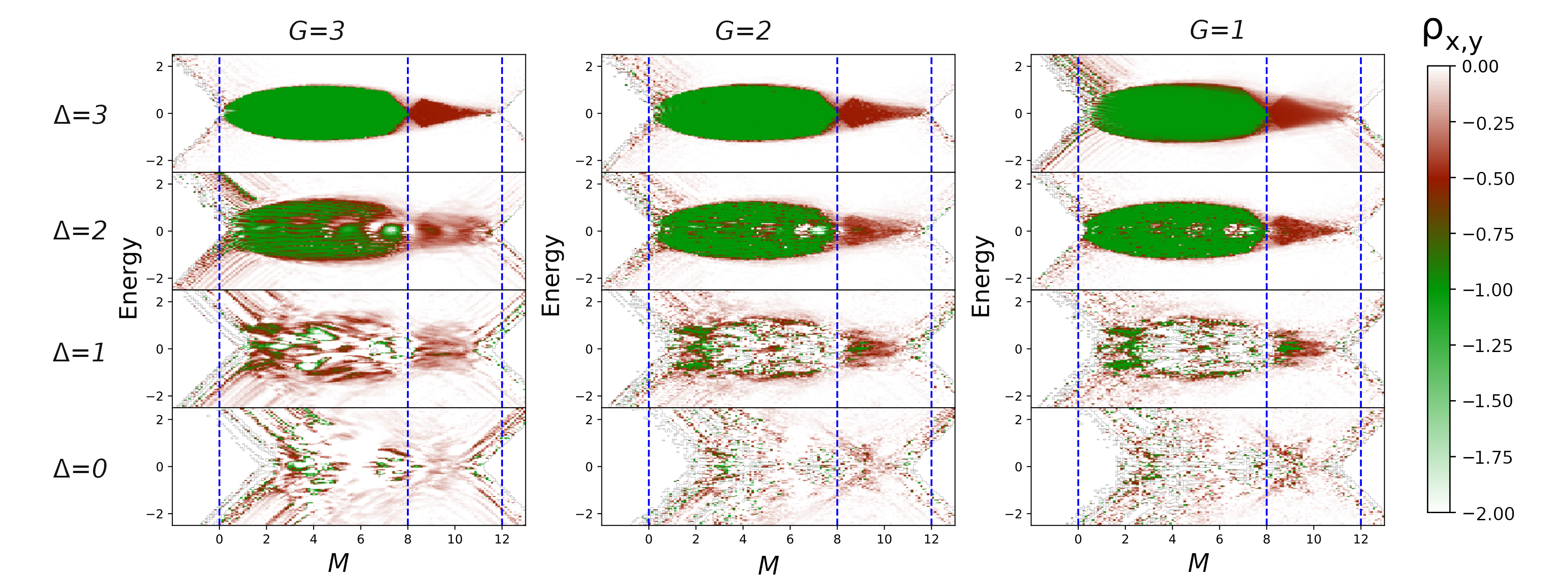}
  \end{center}
    \caption{Hall resistance dependence on $\Delta$ for size generations $G=3,4,5$. The critical value $\Delta=2$ applies for all sizes.}
    \label{fig:allgenerations}
\end{figure*}

A low number of cuts leaves the Hall resistance largely intact.
However, starting from cutting depth $F=3$, the quantization appears less stable and completely vanishes for $F=4$ or $F=5$.
We will discuss later that this is related to the space accessible to a boundary mode. 

This behavior is also observed for other system sizes, as can be seen in Fig.~\ref{fig:allgenerations}.
We find that the relevant measure for the breakdown of the quantization is not the fractal generation $G$ itself but the ``fractal distance'' $\Delta=G-F$.
If $\Delta>2$, the quantization is intact and the Hall resistivity remains unchanged.
For $\Delta<2$, no well-defined region with a finite Hall current exists.
This holds for regions with one edge mode, as well as regions with two edge modes in the non-fractal system.
At $\Delta=2$, there are still regions with a finite Hall resistivity, however they show features of instability.

\begin{figure}
    \centering
    \includegraphics[width=.48\textwidth]{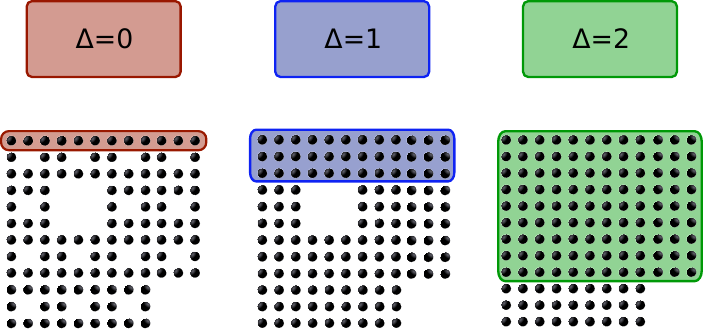}
    \caption{The boundary width $b$ is dependent on the cutting depth (colored regions).
Independent of the size of the fractal, for any given $\Delta$, $b$ remains the same.}
    \label{fig:Delta}
\end{figure}

An interpretation for $\Delta$ can be found by considering the boundary of our system.
Since $\Delta$ refers to the difference in number of actual cuts made compared to the maximum number of cuts possible,
it measures the width of the system boundary that is still intact.
The relation between the number of sites that have not yet been touched by the cutting procedure,  \ie the boundary width $b$ and $\Delta$ is given by
\begin{equation*}
    b=3^\Delta a\;.
\end{equation*}
This relation holds independently of the size of the full system.
A graphical representation for some values of $\Delta$ is given in Fig.~\ref{fig:Delta}.

Since the quantized Hall current is carried by edge modes, details of the edge along which they are traveling should indeed have an impact on their stability.
Understanding why for any boundary with a width of $9$ sites or less ($\Delta\leq 2$) the Hall resistance becomes unstable requires a closer look at the edge states themselves.

\begin{figure}
    \centering
    \includegraphics[width=.48\textwidth]{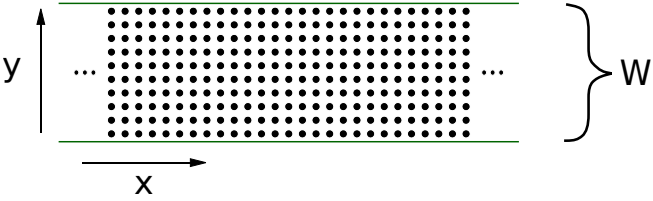}
    \caption{The cylinder setup with width $W$ used for calculating the wave functions of edge modes.
In the $x$-direction the system has periodic boundary conditions.}
    \label{fig:ribbon-setup}
\end{figure}

We consider a cylinder geometry, like the one shown in Fig.~\ref{fig:ribbon-setup},
and investigate the spatial dependence of the wave function transverse to the edge direction. 
By performing a Fourier transform on only the $x$-component, we may study the localization of the edge modes in the finite $y$-direction.
The resulting Hamiltonian is given by
$H_{\text{ribbon}}=\sum_{k} \left [H_{\circ}(k)+H_{\uparrow}(k)+H_{\downarrow}(k) \right]$,
where
\begin{eqnarray}
    \label{eq:ribbonham}
    \begin{split}
    &H_{\circ}(k)=\sum_{y=1}^{W} \psi^{\dagger}_{k,y} \left[\sin (k) \sigma_x + (M-8B+2B\cos( k))\sigma^{z}\right]\psi_{k,y}\;,\\
      &H_{\uparrow}(k)=\sum_{y=1}^W \psi^{\dagger}_{k,y}
      \left[\frac{-\i}{2} \cos (k) \sigma^x + \frac{1}{2}\left(\sin (k) + \i \right)\sigma^y \right.\\
        & \quad\quad\quad\left. \phantom{\frac{-\i}{2}}
        + B\left(1+2\cos (k)\right) \sigma^z\right]\psi_{k,y+1}\;,\\
      &H_{\downarrow}(k)=\sum_{y=1}^W \psi^{\dagger}_{k,y}
      \left[ \frac{\i}{2} \cos (k) \sigma^x + \frac{1}{2}\left(\sin (k)-\i \right) \sigma^y \right.\\
        & \quad\quad\quad \left. \phantom{\frac{-\i}{2}}
        + B\left(1+2 \cos (k)\right) \sigma^z\right]\psi_{k,y-1} \;,
    \end{split}
\end{eqnarray}
and $W$ is the width of the ribbon shown in Fig.~\ref{fig:ribbon-setup}.

\begin{figure}
    \centering
    \includegraphics[width=0.48\textwidth]{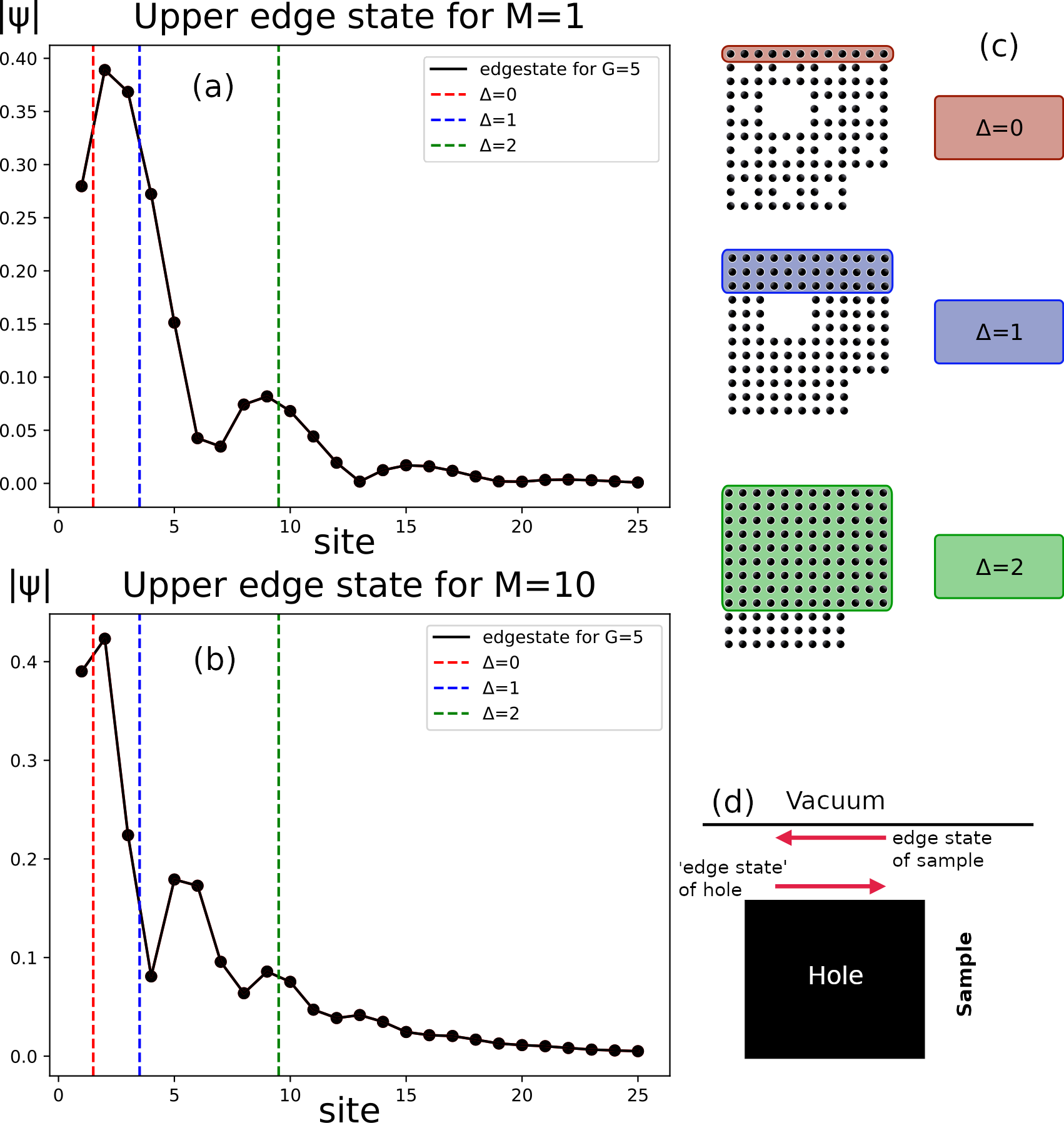}
    \caption{(a) and (b): Wave functions of edge modes for the parameters $M=1$ and $M=10$, and generation $G=5$.
      In the $M=10$ case, (b), there exist two edge modes. The plot shows the zero energy solution at $k=0$.
      Further, the shape of the wave functions agrees perfectly with the solution for $G=4$,
      suggesting that this is the shape also in the infinite system.
(c) Schematic representation of the effects of $\Delta$ on the boundary width.
(d) In the region between the sample edge and the hole in the sample,
there are effectively counterpropagating modes that are not protected from scattering.}
    \label{fig:edgemodes_haldane}
\end{figure}

\begin{figure}
    \centering
    \includegraphics[width=0.7\columnwidth]{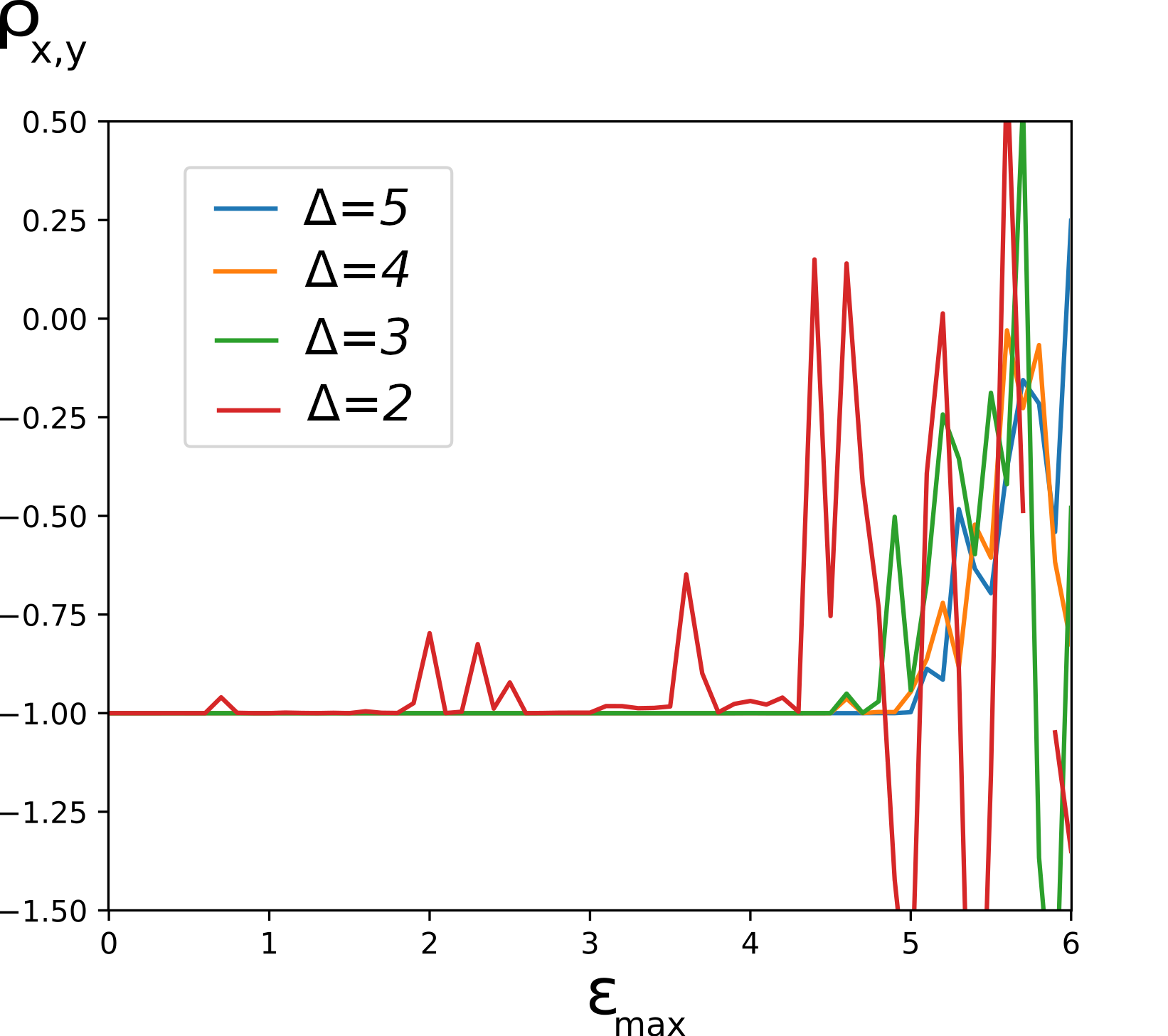}
    \caption{Hall resistance for $M=4$ and $G=5$ in the presence of disorder for SC fractals with various cutting depths.
The critical disorder becomes smaller as the boundary size decreases.}
    \label{fig:disorder_fractal}
\end{figure}

In Fig.~\ref{fig:edgemodes_haldane}, we show the edge states on the non-fractal square lattice for a system size corresponding to generation $G=5$ ($F=0$).
This serves as reference point for the following discussion. 
We show the absolute value of the wave functions as a function of the distance to the edge for the first 25 sites.
There is no visible difference between the shown curve and that of $G=4$,
from which we conclude that for the system sizes considered, finite-size effects are not important. 
We now compare the typical extension to the length scale defined by the cutting procedure.

Vertical lines indicate at which points a cut would effectively terminate the intact boundary of a fractal system for a given $\Delta$.
For $\Delta=0$ ($\Delta=1$) this implies that there are already missing sites in the second (fourth) row  of atoms of the system.
What this means in practice, is that a sizeable weight of the wave function will be localized around holes,
which leads to situations where the different parts of the wave function are effectively counterpropagating, see Fig.~\ref{fig:edgemodes_haldane} (d).

Such systems can therefore not sustain stable edge modes and it is not surprising that the quantization of the Hall conductivity breaks down in these cases, as shown in Fig.~\ref{fig:allgenerations}.
If, however, the boundary is larger than $\Delta=2$, the weight of the wave function is far enough from any holes in the sample. Therefore,
for $\Delta\geq 2$ the Hall resistance in Fig.~\ref{fig:allgenerations} is hardly influenced by the presence of cuts.
The case $\Delta=2$ marks the threshold between both behaviors.
There, the main peak of the wave function still gets supported by the lattice.
However, already the second peak in Fig.~\ref{fig:edgemodes_haldane} (a),(b) cannot completely fit into the boundary when the fractal gets cut out.
This explains why, in the Hall resistance plots, we can still find an area with well defined edge modes,
but in combination with additional substructures that come from the presence of another cut-out edge hybridizing with the outer modes.

Further insight can be gained by considering the stability of edge modes on the fractal under the influence of disorder.
We implement disorder via a random on-site potential $\epsilon$,
which is chosen in the range $\epsilon\in[-\epsilon_{\mathrm{max}},\epsilon_{\mathrm{max}}]$.
The results are shown in Fig.~\ref{fig:disorder_fractal} for the parameter value $M=4$.
Since edge modes were already shown to be unstable for $\Delta<2$, these cases will not be considered.
The results in Fig.~\ref{fig:disorder_fractal} were calculated for $G=5$ fractals, but smaller systems show the same qualitative behavior.
As in the two-dimensional case, \ie $F=0$,
there exists a critical disorder at which edge modes cease to exist.
This critical threshold also exists in the fractal system.
However, the value for such a threshold becomes slightly smaller as $\Delta$ decreases.
At $\Delta=2$, we see some features already for small disorder,
but in general there still seems to be some stable Hall current.
This can be explained by the finite size of the edge modes making them unstable on smaller boundary sizes, as was  previously discussed.

We end this section with a comment.
The reader might wonder if there is a different explanation for the breakdown of topology that does not rely on the mechanism of inner and outer counter-propagating edge modes being gaped out by the presence of the fractal cuts.
For instance, the perturbation introduced by fractal formations could affect different single-particle states differently.
One could then imagine that the periodicity that is introduced by a certain fractal depth happens to be commensurate with electron states that bear the largest Berry curvature $\Omega(k)$.
Those states may then be strongly scattered by the newly added fractal perturbation and may open up trivial gap due to Weiss oscillations\cite{Weiss1991,Weiss1994}.
It is thus justified to wonder what the fate is of the edge modes if one instead of fractal cuts perform periodic cuts of the same size as the cuts at a certain ``fractal-distance'' $\Delta$.

We have not pursued this direction in this work,
but it makes an interesting follow up study.
One may argue that if the hybridization of edge modes is not important,
then topologically non-trivial band structures would persist even at ``periodic'' cuts of $\Delta=1$ or $\Delta=0$, for at least some range of $M$.
One could (but we have not) test this hypothesis by making periodic cuts of finer and finer size, and compare with the fractal cuts. 
The type of ``periodic'' deformations described above would also allow for a multi-band Brillouin zone to still exist, and if the number of bands is reasonably small, one could still compute Chen numbers in the infinite system.

\begin{figure}
  \centering
  \includegraphics[width=0.45\textwidth]{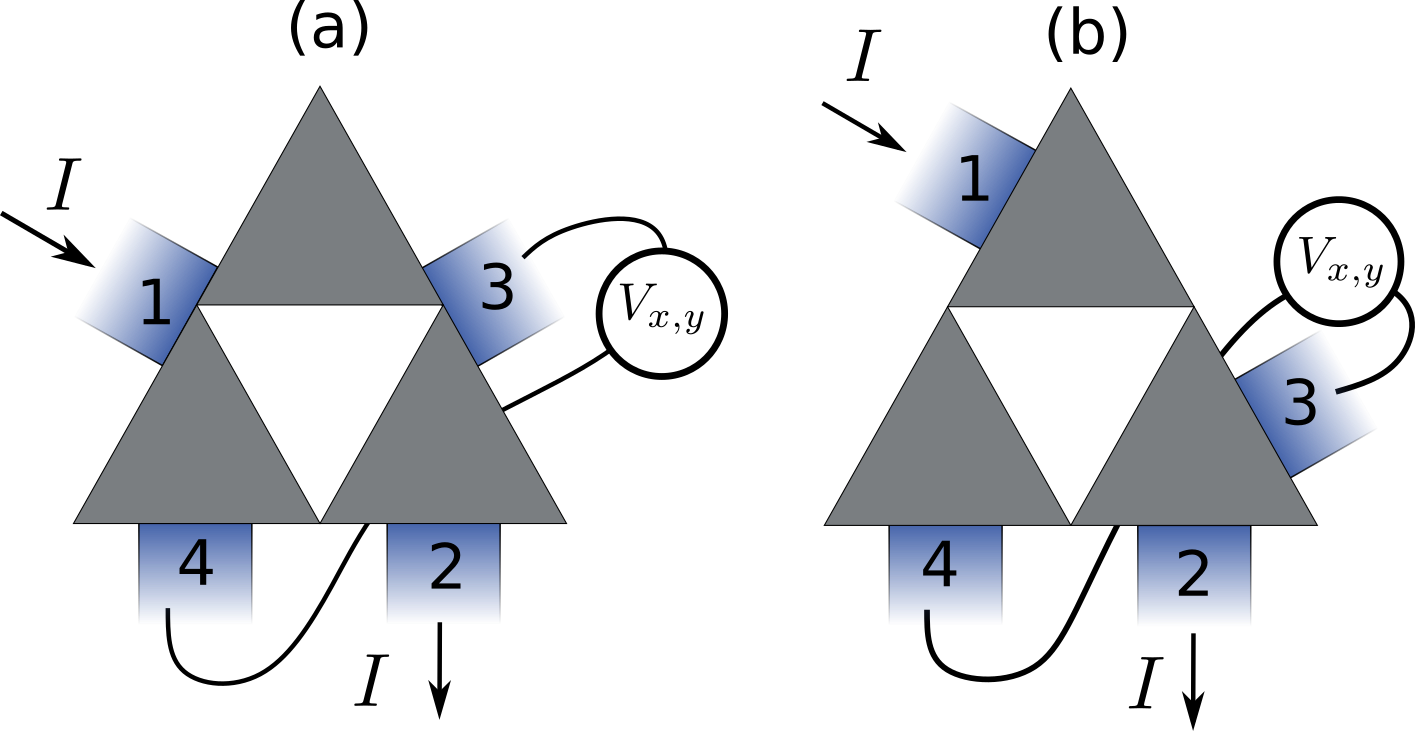}
  \caption{The two different lead setups considered in this work.
The leads are scaled to be 1/4 of the length of the triangle side.
    For purposes of Hall voltage, the setup in (b) yields a more stable reading of $\rho_{xy}$.}
  \label{fig:trianglesetup}
\end{figure}

\begin{figure}
  \centering
  \includegraphics[width=0.49\textwidth]{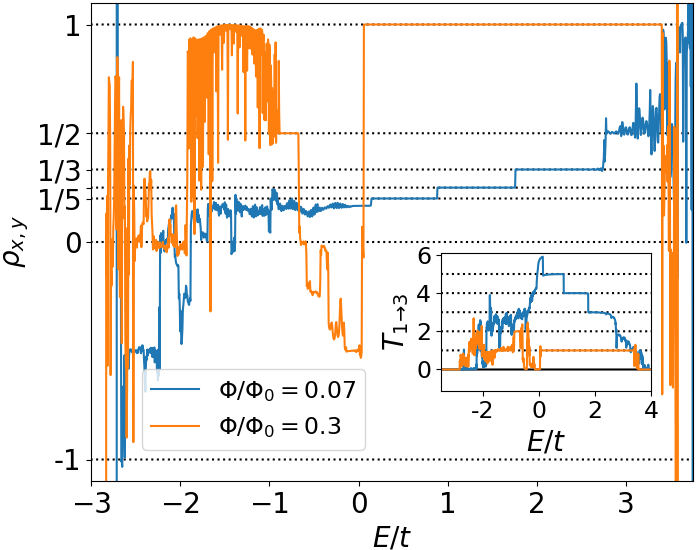}
  \caption{ Transmission and Hall resistivity in a non-fractal triangle for ${\Phi/\Phi_0}=0.07$ and ${\Phi/\Phi_0}=0.3$ with size $G=7$ and with the lead setup shown in Fig.~\ref{fig:trianglesetup} a).\\
    \emph{Main panel}: The Hall resistivity, $\rho_{x,y}$.
    \emph{Inset}: Transmission from lead 1 to lead 3, $T_{1\to3}$.
    For ${\Phi/\Phi_0}=0.07$, bands of Chern numbers up to 5 can be identified,
with the corresponding Hall voltage being quantized at $\rho_{xy} = 1/5 h/e^2$.
    For ${\Phi/\Phi_0}=0.3$, a large energy gap with one edge mode is observed in the energy range $0<E<3.5$.
    \label{fig:HallVoltTriangle}}
\end{figure}

\section{The Sierpinski gasket}\label{sec:gasket}

We return to the Hofstadter model, but on an underlying triangular lattice and the Sierpinski gasket geometry.

Since we are interested in the Hall voltage, this begs the question of how to define the Hall voltage.
In this geometry, we adapt the ``standard'' Hall measurement to the triangular setup by putting two of the leads on the same side.
The measurement then proceeds in the standard way, by running a current between ``opposite'' leads and measuring $\rho_{x,y}$ on the remaining two leads (Fig.~\ref{fig:trianglesetup}).

\begin{figure}
  \centering
  \includegraphics[width=0.49\textwidth]{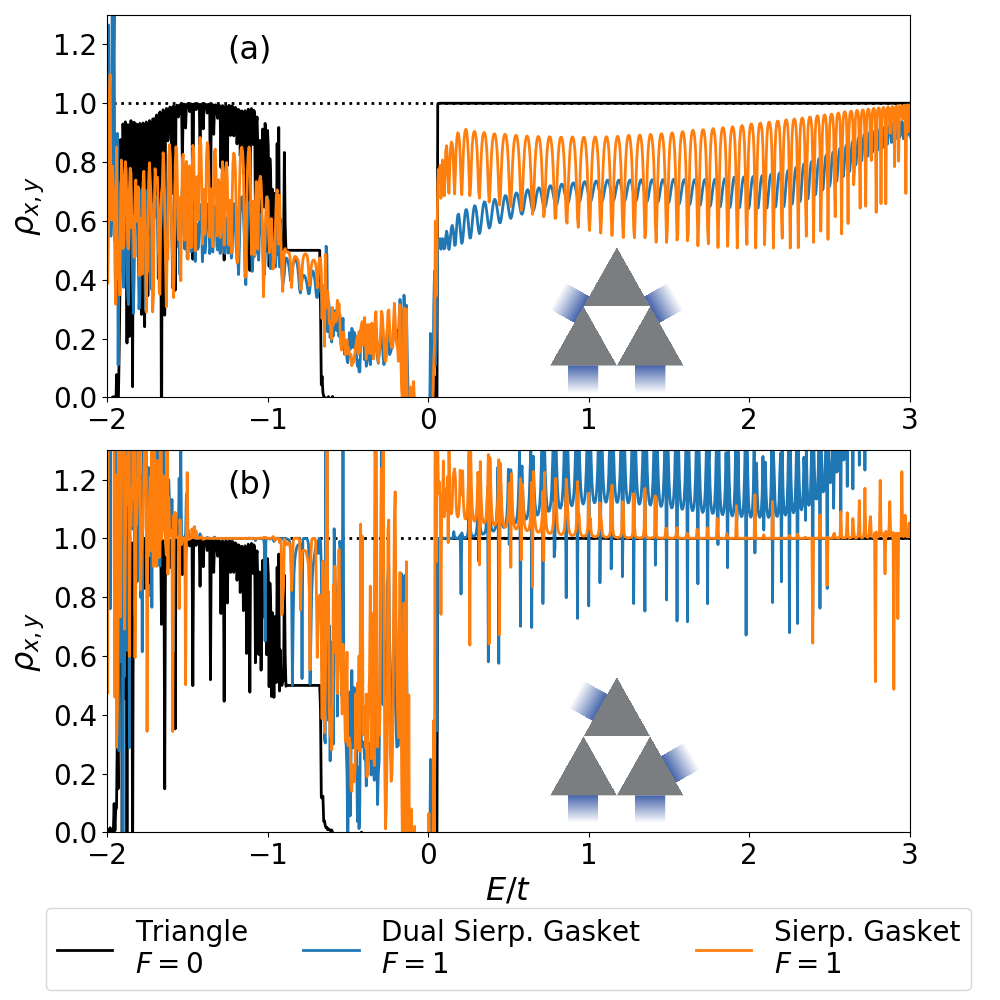}
  \caption{ The Hall voltage for ${\Phi/\Phi_0}=0.3$ with $G=7$ and $F=0$ and $F=1$ with the lead setup shown in a) and b) in Fig.~\ref{fig:trianglesetup}.
    In setup a), $V_HJ$ has trouble picking up any signals of Hall quantization.
    In setup b), some quantization can be seen in $V_H$, but it does depend on the lattice regularization.
    We conclude that the lead setup in Fig.~\ref{fig:trianglesetup} b) is preferable over setup a) for the purposes of detecting a quantized Hall response.
    \label{fig:HallVoltShallow}}
\end{figure}

\subsubsection{Lead placement and shallow cuts}
Here, we explore two different placements of the leads in order to stabilize the Hall voltage measurement, see Fig.~\ref{fig:trianglesetup}.
In setup (a), leads 1 and 3 are placed symmetrically over the pinching points that appear already in the first fractal generation $G=1$.
In setup (b), the leads 1 and 3 are placed asymmetrically next to the $G=1$ pinching points,
but centered on the pinching points for the second fractal generation $G=2$.

As a reference, we study the non-fractal system first.
The results are displayed in Fig.~\ref{fig:HallVoltTriangle}, where we show the Hall voltage at low field ${\Phi/\Phi_0}=0.07$ and high field ${\Phi/\Phi_0}=0.3$.
In the low-field regime, Landau bands with up to Chern number 5 can be identified by the quantization of $\rho_{x,y}$,
whereas in the high-field regime, the spectrum is dominated by the large gap $0<E<3.5$, with only a single edge mode.
Note that in the non-fractal case the two different lead placements in Fig.~\ref{fig:trianglesetup} lead to equivalent results (not shown).

The SG differs substantially from the SC in that already at the shallowest cut $F=1$,
the edge contains pinching points which are only one lattice site wide.
A direct consequence is that these can only allow for one edge mode to pass and this limits the maximal transport through the pinching point to be at most 1.
This also means that we should only expect a quantized Hall voltage that is either $\rho_{xy}=0,\pm1$.

In Fig.~\ref{fig:HallVoltShallow}, we investigate whether there is a preferred way to place the leads to detect this one edge mode.
We here focus on the case ${\Phi/\Phi_0}=0.3$ and compare the $\rho_{xy}$ for both the SG and dual SG regularizations with respect to the placement of the leads in Fig.~\ref{fig:trianglesetup}.
For this purpose, we only make a single cut $F=1$ to try and detect the possibility of a single edge mode.
For the ``symmetric'' placement, we see that $\rho_{x,y}$ shows strong high-frequency fluctuations precisely in the region where a quantized $\rho_{x,y}$ response could be expected.
We speculate the these high frequency oscillations are due to the symmetrically placed leads acting as strong impurities and interfering with the path of the edge mode,
as it tries to navigate the pinching point which acts like a point contact.

On the other hand, in the ``asymmetric'' lead placement (lower panel) a clear Hall plateau is observed for the SG system.
The same can, unfortunately, not be said for the dual SG system in this energy range.
We note however that in the range $-1.5<E<-1$ both the SG and dualSG lattices show a quantized Hall response.
This shows that at least for a shallow cutting, both lattices are able to support edge modes.
We conclude that the ``asymmetric'' placement of the leads is preferred for detecting edge modes, and we will thus only use that one in what follows.

We note that a further consequence of the pinching points at $F=1$ is that ``bulk'' and ``edge'' currents will need to pass though the same point.
This will lead to the possibility of mixing between the transverse and longitudinal resistivity.
Indeed, the oscillatory behavior of the Hall resistivity shown in Fig.~\ref{fig:HallVoltShallow} is similar to the behavior of Hall measurement in 2D electron gasses that occurs when partially filled Landau levels cross the Fermi level as the $B$-field is increased. 
In the current setup, longitudinal resistivity cannot really be measured, as this would require a 6 terminal setup.
Our computational method can easily be modified to include also a 6 terminal setup, but it starts to become computationally expensive, and would introduce another layer of finite-size effects, so we choose not to do it in this work.
Thus, we cannot conclusively say that the longitudinal and transverse resistances are not mixed.

\begin{figure}
  \centering
  \includegraphics[width=0.49\textwidth]{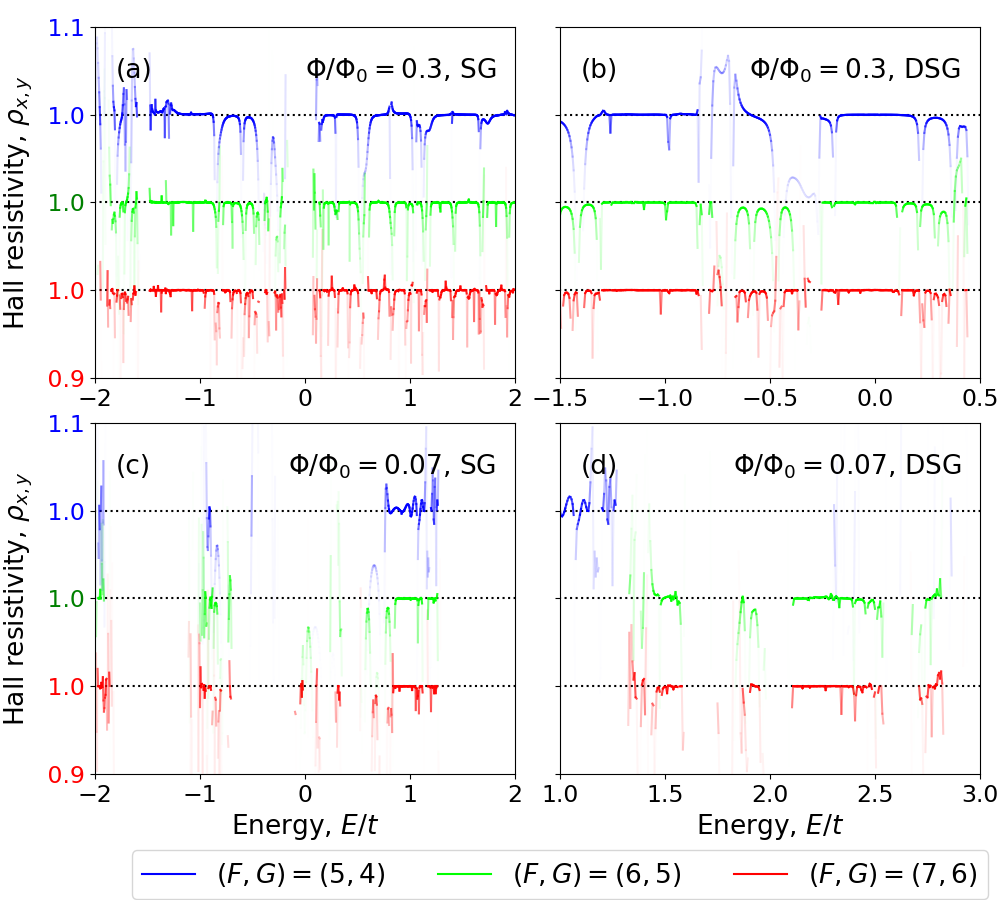}
  \caption{The Hall voltage for ${\Phi/\Phi_0}=0.3$ and ${\Phi/\Phi_0}=0.07$,
and $F=G-1$ for $G=5,6,7$ with the lead setup shown in Fig.~\ref{fig:trianglesetup}(b).
    In each panel, only a segment of the spectrum is shown.
    The value of $\rho_{x,y}$ is shifted by $0.1$ between the three generations for increased readability.
    In addition, values of $\rho_{x,y}$ that deviate from $h/e^2$ are made successively whiter to suppress noise in $\rho_{x,y}$ and to highlight the regions where $\rho_{x,y}$ is quantized. 
    Note that for all three system sizes roughly the same behavior of $\rho_{x,y}$ is observed and that the different system sizes share the same regions with quantized $\rho_{x,y}$.
    It is thus reasonable to suspect that these plateaus will be present also for larger system sizes. 
    \label{fig:HallVoltDeep}}
\end{figure}

\begin{figure}
  \centering
  \includegraphics[width=0.48\textwidth]{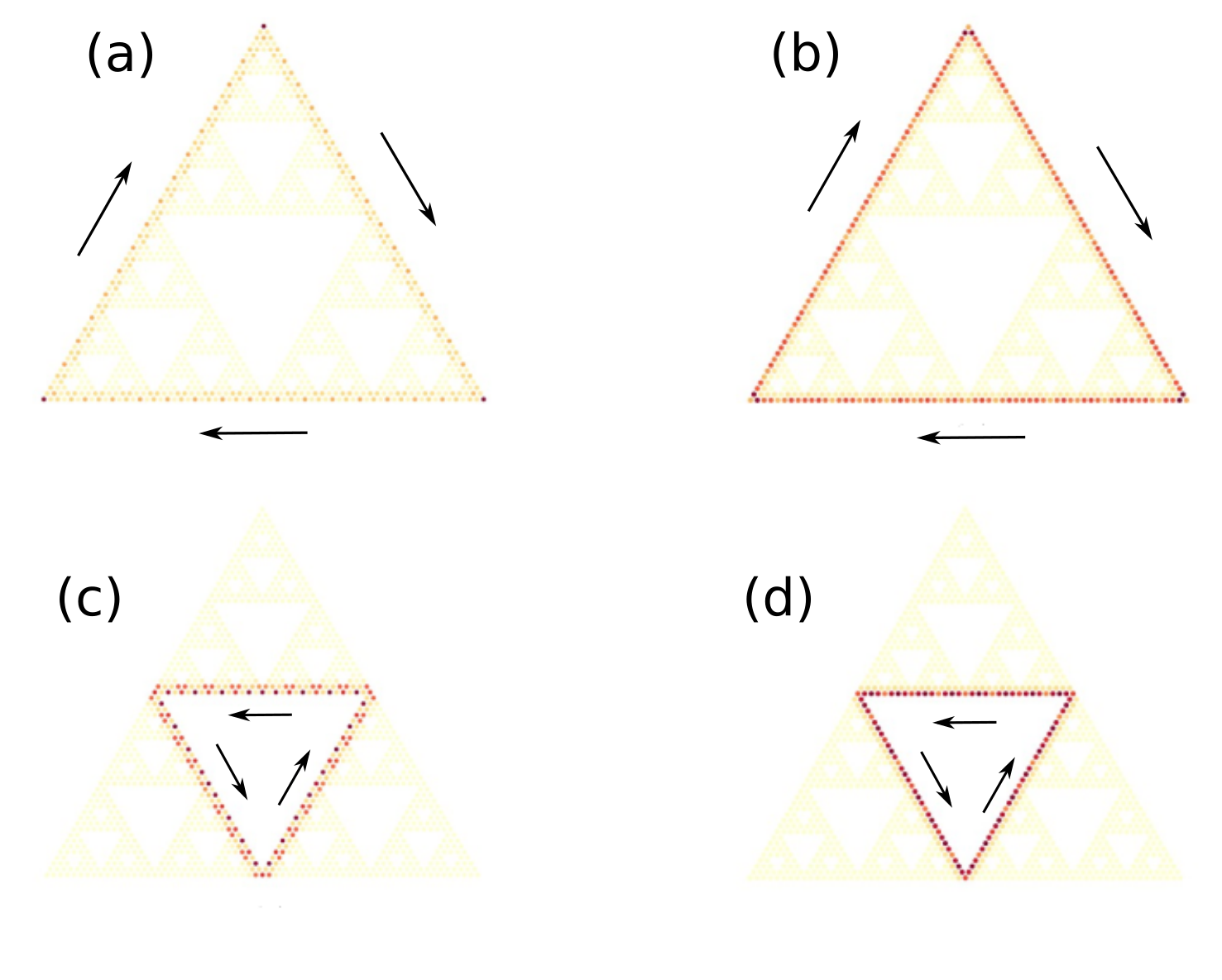}
  \caption{ Eigenstates situated at:
    (a) ${\Phi/\Phi_0}=0.07$, $E = -0.83$,
    (b) ${\Phi/\Phi_0} = 0.33$ , $E = -2.71$,
    (c) ${\Phi/\Phi_0} = 0.07$ , $E = -0.79$,
    (d) ${\Phi/\Phi_0}= 0.33$ , $E = -2.65$.
    In all figures, a disorder $w\epsilon=\in[-0.1,0.1]t$ is used.
    \label{fig:edgemodes_triangle}}
\end{figure}

\begin{figure}
  \centering
  \includegraphics[width=0.48\textwidth]{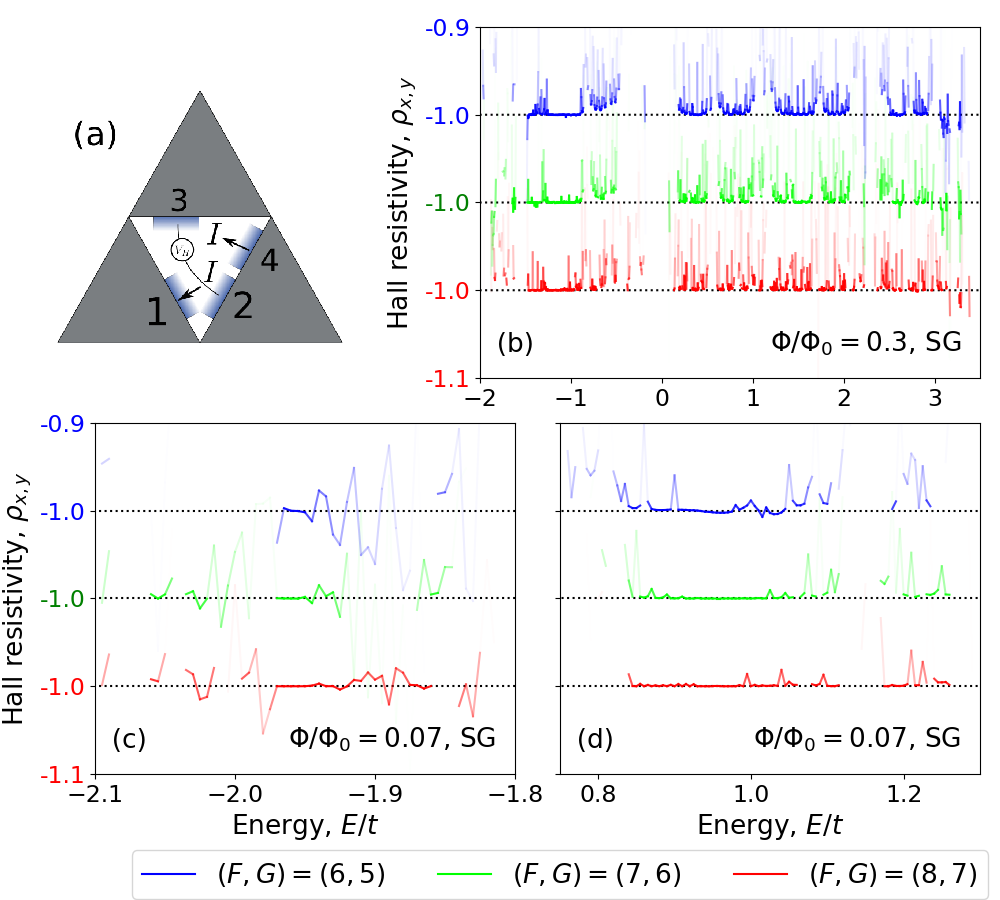}
  \caption{The Hall voltage for ${\Phi/\Phi_0}=0.3$ and ${\Phi/\Phi_0}=0.07$ and $F=G-1$ for $G=5,6,7$, with the lead setup shown in the left panel.
    In each panel, only a segment of the spectrum is shown.
    The value of $\rho_{x,y}$ is shifted by $0.1$, just like in Fig.\ref{fig:HallVoltDeep}.
    Also here, the different system sizes share the same regions with quantized $\rho_{x,y}$.
    Furthermore, the regions with stable Hall resistivity are the same as in Fig.~\ref{fig:HallVoltDeep}.
    \label{fig:HallVoltDeep2}}
\end{figure}

\subsubsection{Fractal cuts}

Next, we perform cuts at maximal depth, such that the fractal depth is $F=G-1$.
The result is displayed in Fig.~\ref{fig:HallVoltDeep} for two different magnetic fields ${\Phi/\Phi_0}=0.07,0.30$ and both the SG and dual SG setup, and $G=5,6,7$.
For increased readability, the curves are shifted relative to each other.
Further, to highlight the regions with quantized Hall resistivity $\rho_{x,y}$, deviations from $\rho_{x,y}=h/e^2$ are fading towards white.

In the panels of Fig.~\ref{fig:HallVoltDeep2}, we can see that for all three system sizes,
roughly the same behavior of $\rho_{x,y}$ is observed.
Also, we note that the different system sizes share the same regions with quantized $\rho_{x,y}$.
It is thus tempting to conclude that the fractal is able to support topological states in the thermodynamic limit.
Indeed, by diagonalizing the Hamiltonian without leads, we find clearly distinguishable edge modes in the full depth fractal.
These are depicted in panels (a) and (b) of Fig.~\ref{fig:edgemodes_triangle}.

\subsubsection{Interior transmission}

Interestingly enough, from the point of view of the pinching points,
there is no particular reason why the edge states of the Sierpinski gasket have to run only on the outside of the gasket.
In fact, the interior of the gasket forms an edge that is just as valid as the exterior edge.
Indeed, in Fig.~\ref{fig:edgemodes_triangle} (c) and (d), one can see edge states at the interior of the gasket.

If we place leads around the inner holes of the main triangle,
as depicted in Fig.~\ref{fig:HallVoltDeep2}, we can clearly see regions of quantized Hall resistivity.
We note two things: (i) the regions of quantized resistivity are the same for the inner and the outer placement of the leads.
(ii) here $\rho_{x,y}=-h/e^2$ instead of $\rho_{x,y}=+h/e^2$, as in the outer placement.

From the second observation, we conclude that for the inner placement the edge modes are propagating as $1\to2\to4\to3\to1$,
whereas in the outer placement the circulation direction is $1\to3\to4\to2\to1$,
as indicated by the arrows in Fig.~\ref{fig:edgemodes_triangle}.

Combining the second observation with the first observation that the interior and exterior quantization happens in the same energy range,
we draw the conclusion that the pinching points of the gasket are not enough to gap out the edge modes.
Rather, it looks as if the pinching point is alternating between sustaining propagation on the exterior and interior edges,
allowing these in practice to coexist for the purposes of transport.

\begin{figure}
  \includegraphics[width=0.40\textwidth]{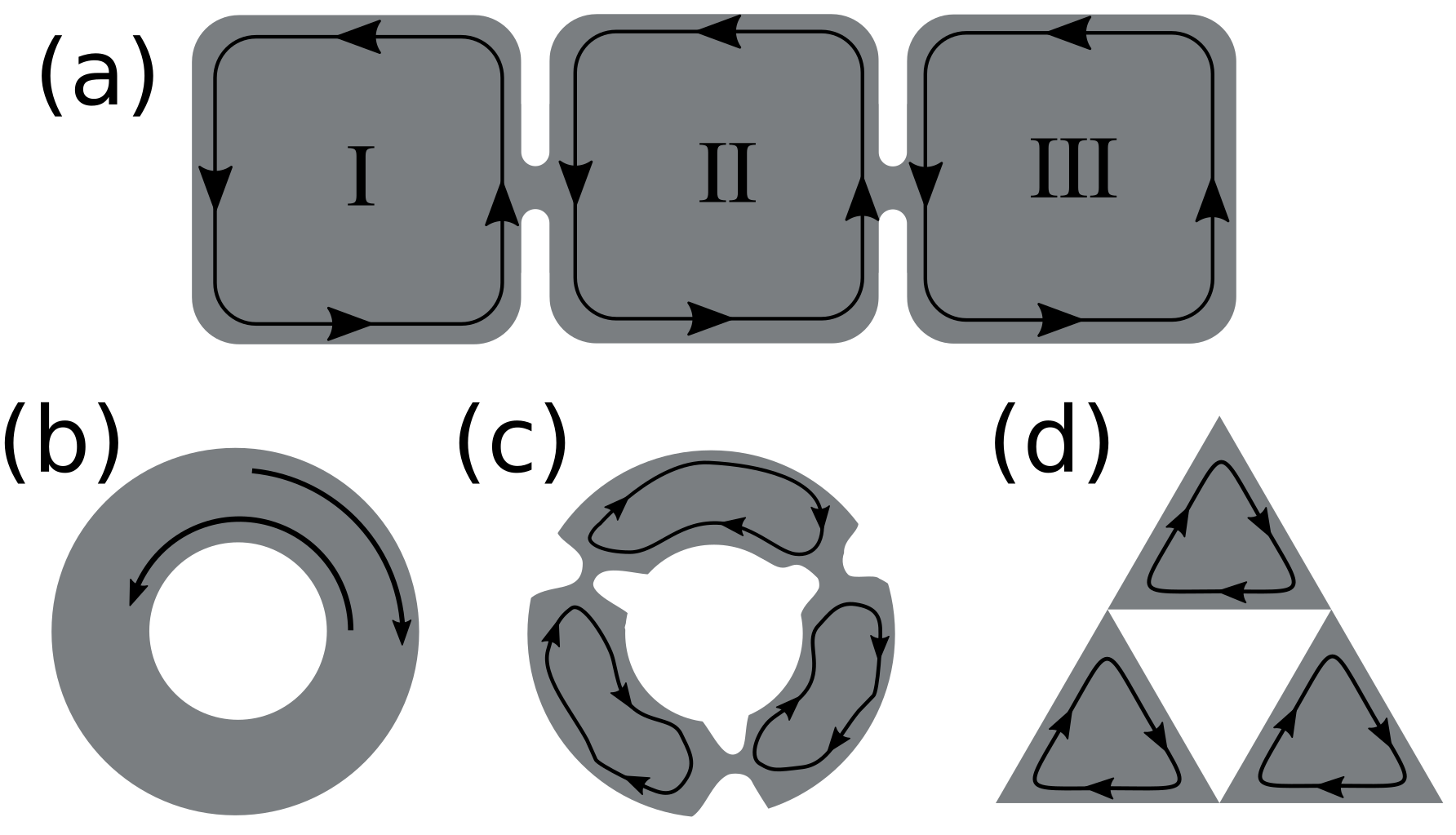}
  \caption{The Sierpinski gasket d) can be likened wit a network of connected quantum Fabry-Perrot interferometers c) or as a pinched Corbino disk.
    In both cases, counter-propagating edge modes need not gap each other out completely.
    \label{fig:FabryPerrot}}
\end{figure}

The observation of coexisting transport can be understood from the perspective of a quantum Fabry-Perrot interferometer, as shown in Fig.~\ref{fig:FabryPerrot}a).
In an interferometer of this kind, edge modes can circulate on each of the regions I-III without being destroyed by the quantum point contacts connecting them. 
At the same time, currents may tunnel between region I and II (or II and III) coherently.
An analogous situation can be found in the case of the Sierpinski gasket with $F=1$.
The three triangles that are formed at $F=1$ can alternatively be thought of as forming a Fabry-Perrot interferometer necklace,
Fig.~\ref{fig:FabryPerrot} (c), or as a pinched Corbino geometry.
In both cases, the counter-propagating edge modes do not gap each other out completely allowing for coexistence and a quantized Hall resistivity response on the inside, as well as the outside boundary.

\section{Conclusion and discussion}\label{sec:conclusion}

In this paper, we investigated the stability of the IQH effect when paradigmatic models are transferred onto lattices that implement fractals with dimensions between one and two, specifically the SC, SG,
and the dual SG.
On general grounds, starting from bulk considerations,
one should not expect a robust quantization of the transverse conductance:
The Chern number is only well defined in two spatial dimensions.

The question can also be investigated from the point of view of protected edge modes.
They are known to exist in two-dimensional IQH systems by virtue of the bulk-boundary correspondence.
We study how their stability is compromised by rendering the lattices increasingly fractal.
We use a spectroscopic method and the Green function method.

Our main finding is that under generic circumstances, the edge channels become unstable and no quantization can be expected.
The main reason for the instability is rooted in the fact that the fractal introduces new edges on the inside of the sample.
The associated 'inner edge states' are counterpropagating with respect to the outside edge states and can hybridize.
This eventually gaps them out. The only exceptions we find are in situations where either the edge states are extremely localized or where the fractal does not extend all the way to the edges due to the cutting depth.
We show that in those situations, the protection against disorder is also strongly reduced compared to the two-dimensional counterpart.

We stress, however, that even though a robust quantization of the transverse conductance is not expected, there are fine tuned situations where it still exists.
In these situations,  many of the topological features that are observed are also scale invariant, in that they only depend on the fractal distance, $\Delta=G-F$, and not on $G$ or $F$ individually.

For future, it would be interesting to investigate other classes of topological insulators on fractal geometries, and also the connection with Weiss oscillations\cite{Weiss1991,Weiss1994} on periodic structures with the same length scale as the fractal cuts.

\section*{Acknowledgements}
C. M. S. acknowledges discussions with Hans Hansson and Vladimir Gritsev.
This work is part of the D-ITP consortium,
a program of the Netherlands Organisation for Scientific Research (NWO) that is funded by the Dutch Ministry of Education, Culture and Science (OCW). C.O.
acknowledges support from a VIDI grant (Project 680-47-543) financed by NWO.

\bibliography{Edge_States_In_Fractals}

\appendix

\section{The Recursive Green Function Method}\label{app:recgf}

In Section~\ref{sec:method}, we introduced the transmission function $T(\omega)$ that depends on the Green functions.
In principle, the retarded or advanced Green function of our system can be calculated by inverting the Hamiltonian of the full system,
\begin{eqnarray}
    \label{eq:Gdef}
    \hat{G}^{R/A}(\omega)&=&\left[\left( \omega \pm i \eta \right) -H\right]^{-1}\;,\nonumber \\
    H&=&H_{\text{Center}}+H_{\text{Leads}}\;,
\end{eqnarray}
where, in the following, whenever not strictly necessary, we will neglect the infinitesimal regulator $\eta \ll 1$.
 Assuming that we are able to integrate out the lead contribution to the full Hamiltonian,
the Green function has the same dimensionality as the center Hamiltonian $H_{\text{Center}}$, \ie it grows with system size.
In our case, if we consider a system of length $L$ and width $W$,
the matrix we need for a full Hamiltonian or Green function will have $\left[N_{\mathrm{orb}}(L+2)W\right]^2$ elements.
The factor $N_{\mathrm{orb}}$ counts the possibility of multiple orbitals on each site, as featured \eg in the Hamiltonian \eqref{eq:realspaceham}.
Having $L+2$ sites in one direction instead of only $L$ is a result of adding one layer of sites representing the leads on either side.
This means that using exact diagonalization,
by directly solving Eq.~\eqref{eq:Gdef},
in order to calculate the Green functions is not feasible for larger systems.

\begin{figure}
    \centering
    \includegraphics[width=.45\textwidth]{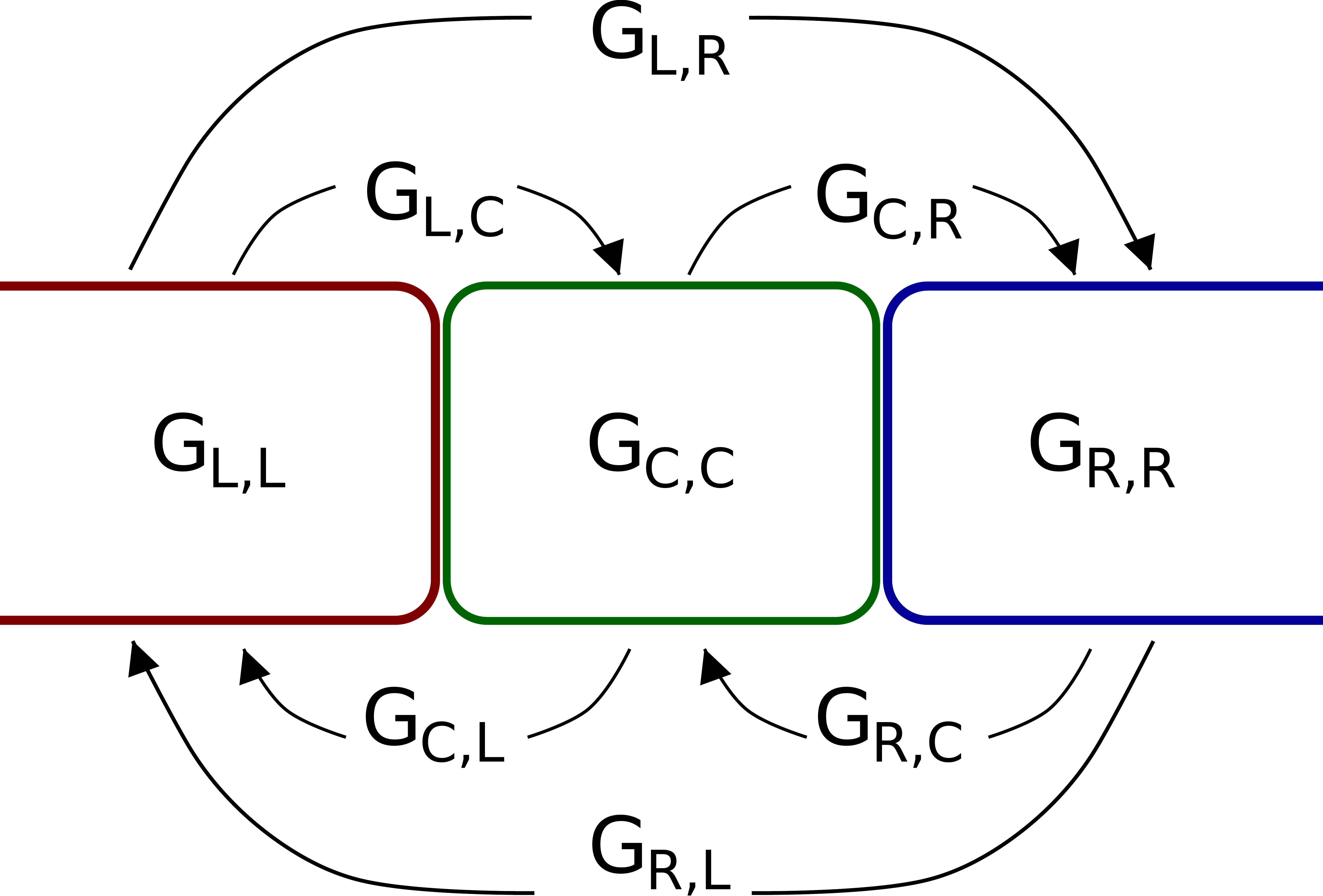}
    \caption{Visualization of the propagation that Green function describes.}
    \label{fig:greenfunctions}
\end{figure}

\begin{figure*}
    \centering
    \includegraphics[width=0.70\textwidth]{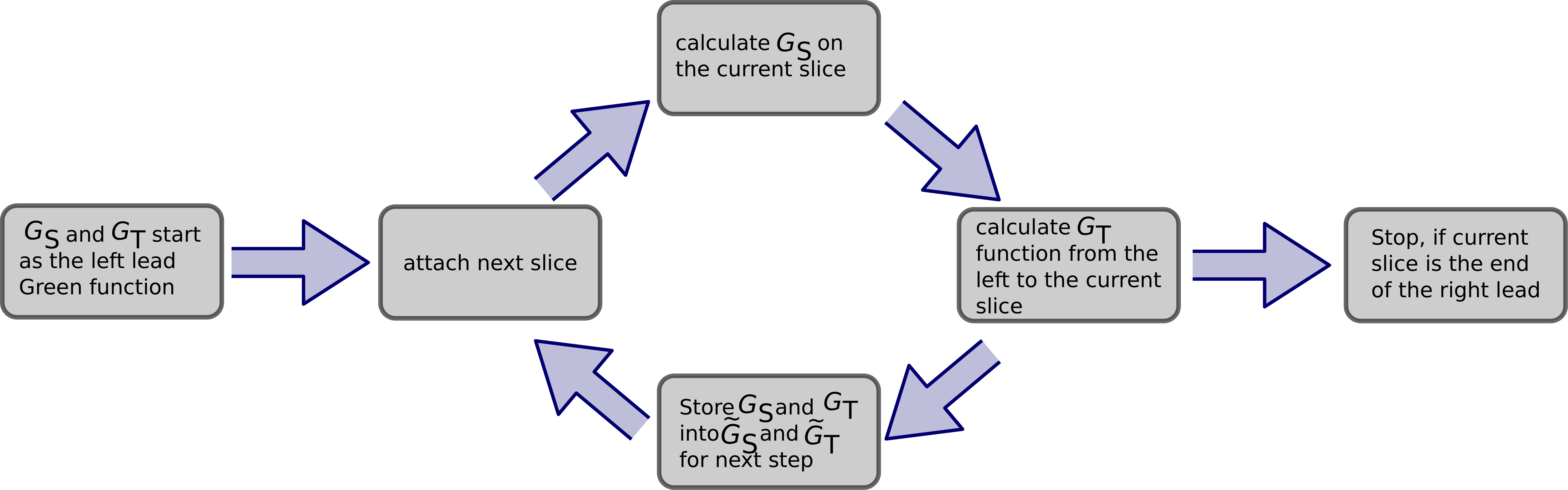}
    \caption{A flowchart depicting the steps taken in the recursive Green function algorithm.
      Each step is described in the text in Sec.~\ref{app:recgf}.}
    \label{fig:flowchart}
\end{figure*}

Fortunately, the transmission function Eq.~\eqref{eq:transmission} does not depend on knowing the full Green function matrix.
Instead, it is enough to know the part of the Green function responsible for propagation from left to right together with the lead contributions.

In order to visualize the dimensionalities of different parts of the Green function, we can write it as a matrix of matrices
\begin{eqnarray*}
    \hat{G}(\omega)&=&\left( \begin{array} {ccc}G_{L,L} & G_{L,C} & G_{L,R} \\
                                     G_{C,L} & G_{C,C} & G_{C,R}\\
                                    G_{R,L} & G_{R,C} & G_{R,R}
                    \end{array} \right) \nonumber\\
        &=& N_{\mathrm{orb}}^2
                    \left(\begin{array}{ccc}
                        W\times W & W \times LW & W \times W   \\
                        LW \times W & LW\times LW & LW \times W \\
                        W\times W & W \times LW & W \times W   \\
                    \end{array}\right) \;,
\end{eqnarray*}
the interpretation of which is depicted in Fig.~\ref{fig:greenfunctions}.
Note, that while the amount of matrix elements in the whole system grows with the volume of the scattering area, the corner elements $G_{L,R}$,
$G_{L,R}$ that describe propagation from one lead to another only grow as the surface of the leads attached on the sides.
The same holds for the other corner elements $G_{R,R}$ and $G_{L,L}$,
which will be required to compute the lead contributions $\Gamma_L$ and $\Gamma_R$ in Eq.~\eqref{eq:Broadening}.
Therefore, if there is a way to compute them without solving the whole system,
we will be able to compute the transmission function \eqref{eq:transmission} for much larger system sizes.
In the following, we will show that this is indeed achievable using a recursive approach, a flowchart of which is shown in Fig.~\ref{fig:flowchart}.

Since we are interested in the transport from the left to the right lead,
the starting and ending point of the recursion are always given by the leads themselves.
Thus, the first Green function to be computed is always the surface Green function of the left lead $G_L$, before it is attached to the scattering region $S$.
For a semi-infinite lead described by \eg a simple tight-binding chain, $G_L$ can even be solved exactly,
but in more complicated situations $G_L$ needs to be determined numerically.

\begin{figure*}
  \begin{center}
    \includegraphics[width=0.8\textwidth]{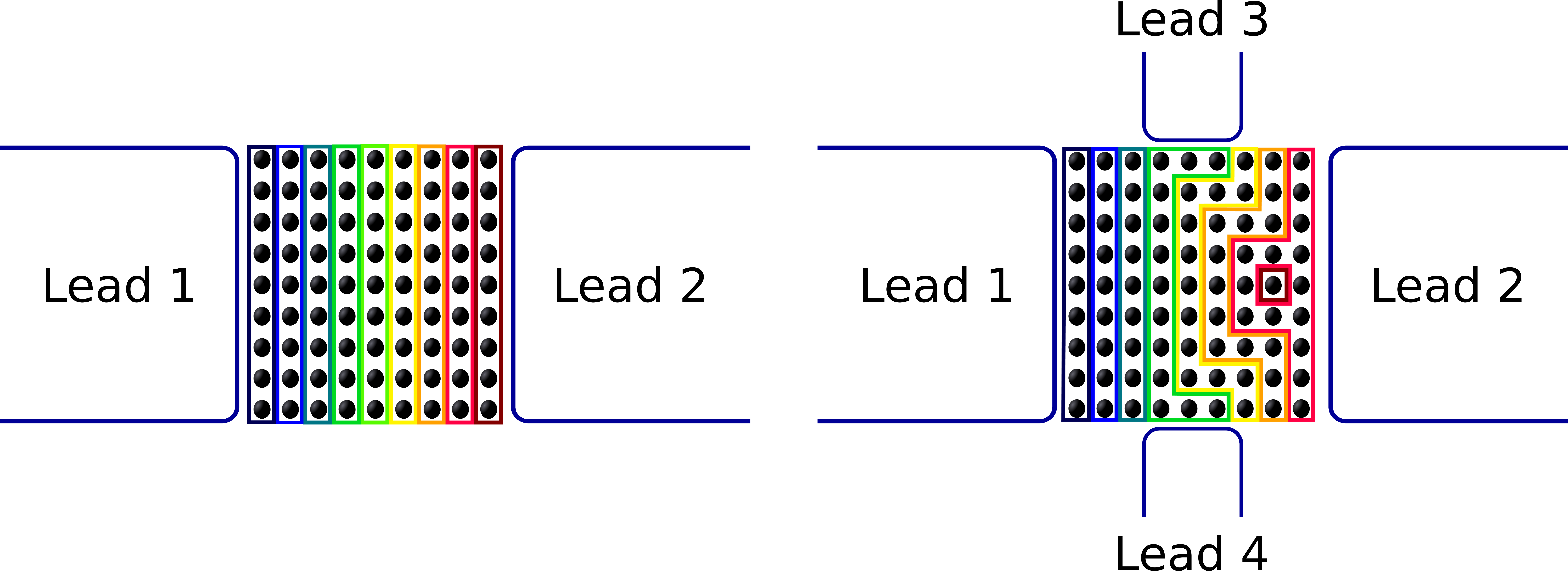}
  \end{center}
  \caption{Slicing procedure for a two-lead setup (left) compared to a four-lead setup (right). Each color indicates one step in the slicing procedure.}
  \label{fig:multislice}
\end{figure*}

Each slice contains the sites to which the system from the previous step directly connects, and that were not considered before.
In a simple square geometry, this implies that the cuts are equivalent to taking all sites that share the $x$-coordinate $x=j+1$ if the previous slice ended at $x=j$.
This remains true even if a fractal is cut into that square geometry.
An example of a simple linear slicing is shown in the left panel of Fig.~\ref{fig:multislice}.

For each slice, interaction matrices $\hat{V}_{\leftarrow}$ and $\hat{V}_{\rightarrow}$ are determined by taking the appropriate terms out of the Hamiltonian that connects both slices.
For the Haldane model in \eqref{eq:realspaceham}, this implies that if we attach a slice that connects sites with $x$-coordinate $j+1$ to the previous slice,
where $x=j$, the interaction matrices are symbolically given by
\begin{eqnarray}
    \hat{V}_{\rightarrow}&=&H_{\rightarrow}+H_{\nearrow}+ H_{\searrow}\;,\nonumber \\
    \hat{V}_{\leftarrow}&=&H_{\leftarrow}+H_{\nwarrow}+H_{\swarrow}\;, 
\end{eqnarray}
where it is understood that only terms containing elements on both collumn $j$ and $j+1$ and included.
Note that on a fractal geometry, some of these Hamiltonian elements may not exist due to missing lattice sites.
This implies that the amount of sites that connect to a slice, as well as the size of a slice itself are not constant,
and therefore the dimensions of the interaction terms and Green functions vary.

If we let $G^{(n)}_{j,k}$ denote the Green's function from slice $j$ into slice $k$ after $n$ slices have been attached,
and $H_n$ denote the Hamiltonian of the $n$th slice, the recursive algorithm reads
\begin{eqnarray}
  G_S=G^{(n)}_{n,n}&=&\left(\omega - H_{n}-V^n_{\leftarrow}G^{(n-1)}_{n-1,n-1} V^n_{\rightarrow}\right)^{-1}\;,\nonumber\\
  G_T=G^{(n)}_{0,n}&=&G^{(n-1)}_{0,n-1}V^n_{\rightarrow}G^{(n)}_{n,n}\;.
     \label{eq:recursion}
\end{eqnarray}
We may take $G^{(0)}_{0,0}=G_{\text{Lead}}$ to denote the initial semi-infinite lead.
Note how $G_S$ and $G_T$ get updated as more slices are added to the system, whereas $G^{(n)}_{j,k}$ refers to a specific setup.

In the last step of the algorithm $n=N+1$, the right lead needs to be attached to the scattering area.
In this step, the bare Green function of the new slice gets replaced by the lead Green function again, such that 
\begin{eqnarray*}
G_{R,R}&=&G^{(N+1)}_{N+1,N+1}=\left( G_{\text{Lead}}^{-1}-V^{N+1}_{\leftarrow}G^{(N)}_{N,N} V^{N+1}_{\rightarrow}\right)^{-1},\nonumber\\
G_{L,R}&=&G^{(N+1)}_{0,N+1}=G^{(N)}_{0,N}V^{N+1}_{\rightarrow}G^{(N+1)}_{N,N}\;.
\end{eqnarray*}

Note that the transmission~\eqref{eq:transmission} as well as the recursive algorithm~\eqref{eq:recursion} require the lead Green's function $G_{\text{Lead}}$.
Since it is needed as a starting point for the recursive algorithm, it has to be found before applying the recursion method.
However, computing it is closely related to finding the surface function $G^{(n)}_{n,n}$ in the recursion Eq.~\eqref{eq:recursion} with the addition that, for a semi-infinite system,
adding another slice does not change the surface Green function.
This implies that we need to solve the self-consistency equation
\begin{equation*}
    \label{eq:Glead}
        G_{\text{Lead}}=\left(\omega -H_S-V_{\leftarrow}G_{\text{Lead}} V_{\rightarrow}\right)^{-1}\;,
\end{equation*}
where here, $H_S$ refers to the Hamiltonian on the surface of a semi-infinite lead.
In this work, the leads are assumed to be simple tight-binding chains with nearest-neighbor interaction,
which in the case of Haldane model do not couple the two orbitals of the scattering area.
Thus here, $H_S$ is a one-dimensional tight binding chain with a length equal to the width of the scattering area.

\section{Recursive Greens function slicing for four leads}\label{app:slicing}

In this work, we are studying a scattering area with more than two leads.
As a result, the slicing procedure that was discussed briefly in Appendix \ref{app:recgf} needs to be adjusted.

In the treatment of the leads, the direction in which they extend infinitely was integrated out exactly,
such that they can be coupled to the scattering area as an effective self-energy.
On the inverse Green function level, this self-energy affects the sites that directly couple to the leads only.
Thus, in order for the recursive method to work, we need to be careful when we add a slice of our scattering area that is connected to an external lead.
To be precise, we need to slice our system such that each lead is attached to one, and only one, slice.
In Fig.~\ref{fig:multislice} we give an example for a slicing procedure in a four-lead setup,
in direct comparison to the slicing of the same scattering area for a two-lead system.

Furthermore, the simple fact that we consider currents between more than two leads, as shown in Fig.~\ref{fig:fourterminal},
means that the target sites for which we calculate the transmission Green function will not always sit in the very last slice.
Thus, when a slice labeled by index $n$ is attached, in addition to the computation of the surface Green's function $G_S=G^{(n)}_{n,n}$ and transmission Green function $G_T=G^{(n)}_{n,\lambda}$ from the left lead to the right,
one must now keep track of one $G_T^{(\lambda)}$ for each slice that is attached.

For this purpose, we now distinguish $V^n_\to$, which labels hopping from slice $n-1$ to slice $n$,
and $V_{\lambda,n}$, which labels hopping from lead $\lambda$ to slice $n$.
Just like $V^n_\to\propto\delta_{m-1,n}$ also $V_{\lambda,n}\propto\delta_{n_\lambda,n}$ is nonzero only for one specific $n=n_\lambda$.

When new leads are attached together with a slice $n$, the surface Green function $G_S$ gets modified with the Green functions for the new leads as 
\begin{widetext}
\begin{align}
  \begin{split}\label{eq:recursion-new-1}
  G^{(n)}_{n,n} &= \left(\omega - H_{n}
  -V^n_{\leftarrow}G^{(n-1)}_{n-1,n-1} V^n_{\to}
  -\sum_{\lambda,n_\lambda=n}V_{n,\lambda}G^{(\lambda)}_{\text{Lead}} V_{\lambda,n}
  \right)^{-1}\;.
  \end{split}
\end{align}

The update of the transmission Green function from the leads to the latest slice then depends on whether the lead was added at step $n$ or previously,
\begin{align}
  \begin{split}\label{eq:recursion-new-2}
    G^{(n)}_{n,\lambda} &= \begin{cases}
      G^{(n)}_{n,n} V^n_{\leftarrow} G^{(n-1)}_{n-1,\lambda} &,\;n>n_\lambda  \;,\\
      G^{(n)}_{n,n} V_{n,\lambda} G^{(\lambda)}_{\text{Lead}} &,\;n=n_\lambda  \;, \\
      G^{(\lambda)}_{\text{Lead}} &,\;n>n_\lambda\;.
    \end{cases}
  \end{split}
\end{align}

Finally, the transmission Green's function between the leads is also updated as follows:

\begin{align}
  \begin{split}\label{eq:recursion-new-3}
    G^{(n)}_{\lambda_1,\lambda_2} &= \begin{cases}
      G^{(n-1)}_{\lambda_1,n-1} V^n_\to G^{(n)}_{n,\lambda_2}+G^{(n-1)}_{\lambda_1,\lambda_2}
      &,\;n_{\lambda_1},n_{\lambda_2}<n\\
      G^{(n-1)}_{\lambda_1,n-1} V^n_\to G^{(n)}_{n,\lambda_2}
      &,\;n_{\lambda_1}<n=n_{\lambda_2}\\
      G^{(\lambda_1)}_{\text{Lead}} V_{\lambda_1,n} G^{(n)}_{n,\lambda_2}
      &,\;n_{\lambda_1}=n>n_{\lambda_2}\\
      G^{(\lambda_1)}_{\text{Lead}} V_{\lambda_1,n} G^{(n)}_{n,n} V_{n,\lambda_2} G^{(\lambda_2)}_{\text{Lead}}
      &,\;n_{\lambda_1}=n=n_{\lambda_2}\;.
    \end{cases}
  \end{split}
\end{align}

\end{widetext}

In this notation, the upper index $n$ again refers to the recursion step,
\ie the slice that gets attached.

After the final slice $n=N$, \ie after lead $2$ is attached in Fig.~\ref{fig:multislice},
the lead transmission function in Eq.~\eqref{eq:transmission} can be computed directly from Eq.~\eqref{eq:recursion-new-3},
by identifying $G^R_{\lambda_1,\lambda_2}=G^{(N)}_{\lambda_1,\lambda_2}$ and $G^A_{\lambda_1,\lambda_2}=G^{(N)}_{\lambda_2,\lambda_1}$.

\end{document}